\documentstyle[12pt,aaspp4]{article}

\begin{document}

\def\stara{1E 2259$+$586}
\def\starb{1E 1048$-$593}
\def\starc{4U 0142$+$614}
\def\stard{RXJ 1838$-$0301}
\def\be{\begin{equation}}
\def\ee{\end{equation}}
\def\lsim{\lesssim}
\def\gsim{\gtrsim}
\def\gta{\gtrsim}
\def\lta{\lesssim}
\def\v{{\vec v}}
\def\B{{\vec B}}
\def\del{{\vec\nabla}}
\def\u{\vec u}
\def\k{\vec k}
\def\eps{\epsilon}
\def\kb{k_B}
\def\eb{\epsilon_b}
\def\by{{Y_l}}
\def\yn{{(1-Y_e)}}

\title{TRANSPORT OF MAGNETIC FIELDS IN\break
CONVECTIVE, ACCRETING SUPERNOVA CORES}

\author{Christopher Thompson}
\affil{Physics and Astronomy, University of North Carolina CB3255, 
Chapel Hill, NC 27599;\hfil\break
Canadian Institute for Theoretical Astrophysics, Toronto, ON
M5S 3H8}
\author{Norman Murray}
\affil{Canadian Institute for Theoretical Astrophysics, Toronto, ON
M5S 3H8}

\begin{abstract}
We consider the amplification and transport of a magnetic field in
the collapsed core of a massive star, including both the region between the
neutrinosphere and the shock, and the central, opaque core.  An analytical
argument explains why rapid convective overturns persist within
a newly formed neutron star for roughly 10 seconds ($> 10^3$ overturns),
consistent with recent numerical models.  A dynamical balance between
turbulent and magnetic stresses within this convective layer corresponds
to flux densities in excess of $10^{15}$G.  Material
accreting onto the core is heated by neutrinos and also becomes strongly
convective.  We compare the expected magnetic stresses in this
convective `gain layer' with those deep inside the neutron core.

Buoyant motions of magnetized fluid are greatly aided by the intense 
neutrino flux.  We calculate the transport rate through a medium
containing free neutrons, protons, and electrons, in the limiting cases of
degenerate or non-degenerate nucleons.  Fields stronger than $\sim 10^{13}$
G are able to rise through the outer degenerate layers of the neutron 
core during the last stages of Kelvin-Helmholtz cooling (up to 10 seconds 
post-collapse), even though these layers have become stable to convection.
We also find the equilibrium shape of a thin magnetic flux rope in the dense 
hydrostatic atmosphere of the neutron star, along with the critical 
separation of the footpoints above which the rope undergoes 
unlimited expansion against gravity.  The implications of these results 
for pulsar magnetism are summarized, and applied to the case of 
late fallback over the first $10^3-10^4$ s of the life of a neutron star.
\end{abstract}
\vskip .3in
\centerline{\it Astrophysical Journal, in press}

\section{Introduction}\label{sa}

Young neutron stars are believed to be strongly magnetized,
based on measurements of their spin behavior (\cite{kulkarni}, and
references therein) and the polarization of their pulsed emissions
(\cite{ML}).  The inferred dipole fields are in the range
$10^{11}-10^{14}$G. In a broader sense, neutron star magnetic fields
of this magnitude are weak.  Consider, for example, a newly formed neutron
star within which neutrinos of all types are temporarily trapped:
near the neutrino photosphere (neutrinosphere) this dipole field 
makes only a tiny contribution to the hydrostatic stresses,
$B_{\rm dipole}^2/8\pi P \sim 10^{-9}(B_{\rm dipole}/10^{12}~{\rm G})^2$,
as compared with $\sim 10^{-6}$ in the Sun.  Over the last
several years there has been a growing accumulation of evidence
pointing to the existence of neutron stars in which the magnetic
fields are so strong as to be transported rapidly throughout the core
and deep crust of the star, even at the remarkably short age of $\sim
10^4$ yr.  This evidence has come in several forms: the short but
dramatically bright outbursts of the Soft Gamma Repeaters; the
persistent, pulsed output of X-rays from these sources and the
non-bursting Anomalous X-ray pulsars; and the rapid braking of the
rotation of the SGRs and AXPs.  These developments are exciting, in
part, because direct physical measures of strong magnetic fields in
isolated neutron stars such as radio pulsars have been difficult to
obtain.

The physical origin of magnetism in neutron stars is difficult to pin
down because there are few direct observational probes of
the evolving cores of massive stars.  Nonetheless,
a simple physical argument points directly to the violent convective
motions that develop in the supernova core as the predominant source
of free energy.  The kinetic energy carried by this convection (both
inside the neutrinosphere of the nascent neutron star, and between
the neutrinosphere and the shock) is $\sim 100$ times greater relative to the
gravitational binding energy than it is during any convective episode
driven by nuclear burning (\cite{TD93}, hereafter TD93).  
From this perspective, the strength
of the seed magnetic field (present, e.g., at the end of the main sequence
evolution) becomes of secondary consequence to the final disposition of 
magnetic fields in the neutron star.  There is growing empirical evidence
-- from chromospheric and coronal activity in fully convective stars
(Delfosse et al. 1998; \cite{kr})  and from the small scale,
intranetwork magnetic field of the Sun (e.g. \cite{DYR}) --
that magnetic and turbulent stresses reach a dynamical balance in
a fluid of very high magnetic Reynolds number, even if that fluid is
slowly rotating.

In this paper, we consider afresh the transport of magnetic fields in 
the collapsed core of a supernova. We focus in particular on 
the role of the intense neutrino flux in facilitating transport
of magnetic fields across convectively stable layers that would
otherwise bury dynamo-generated magnetic fields.  Before the supernova
shock escapes to large radius, the neutrinosphere is enveloped by a 
fairly thick layer of material ($\Delta M \sim 0.01$-$0.1 M_\odot$) which 
is stabilized to convection by a gradient in electron fraction $Y_e$ 
(Keil, Janka \& Mueller 1996, hereafter \cite{KJM}).  Even as the
shock propagates outward, a reverse shock forms that can deposit
up to $\sim 0.1\,M_\odot$ onto the neutron core over $\sim 10^3$ s
(\cite{ww95}; Fryer, Colgate, \& Pinto 1999, and references therein).  

The plan of this paper is as follows.  Section 2 and the Appendix give a fresh
analysis of persistent convection inside the neutrinosphere of a newly
formed neutron star.  We focus on the forcing of convection by
secular neutrino cooling, when the lepton number gradient begins 
to stabilize the star but a negative entropy gradient persists.  We connect
our calculation to those of Lattimer \& Mazurek (1981) and Reisenegger 
\& Goldreich (1992), who calculate the Brunt-V\"ais\"al\"a frequency 
in hot and cold neutron stars.  We also note that
continuing fallback onto the neutron core will induce rapid convection
outside the neutrinosphere that can deposit a strong (but tangled)
magnetic field in the surface layers of the star.  

The rate of radial
transport of a magnetic field in and around the forming neutron
core is calculated in detail in Section \ref{transport}.   Turbulent
pumping moves magnetic flux outward rapidly through the convective
core, in the direction of decreasing turbulent diffusivity (\S \ref{pumping})
Outside the neutrinosphere, neutrino
absorption increases the entropy of a magnetic flux rope with respect
to its surroundings and allows it to move outward through a stably
stratified atmosphere (\S \ref{fivetwo}). At late times, when the
optically thick interior of the star is no longer convective,
neutrinos induce transformations between protons and neutrons that
allow a strongly magnetized parcel of fluid to move across a
stablizing lepton number gradient (\S \ref{latet}).  The net result is
that magnetic fields stronger than $10^{13}-10^{14}$ G will rise
buoyantly to the surface of a newly formed neutron star.  The
equilibrium configuration of a magnetic flux rope that extends outward
from the convective interior of the star through the dense outer
atmosphere is addressed in Section 4.  

We summarize in \S 5 the implications for pulsar magnetism,
discussing the relative importance of the interior and exterior
dynamos, and applying our results to the case of late fallback
(over $\sim 10^3-10^4$ s).  The gravitational binding energy of the
accreting material is converted to electron-type neutrinos, whose luminosity 
$G\dot M M_{\rm NS}/R_{NS} = 7\times 10^{49}\,
(\dot M/10^{-4}\,M_\odot {\rm s^{-1}})\,(M_{\rm NS}/1.4\,M_\odot)\,
(R_{\rm NS}/10~{\rm km})^{-1}$ erg s$^{-1}$ is smaller, by about two 
orders of magnitude, than the luminosity emitted during the prompt 
$\sim 30$ s Kelvin-Helmholtz phase.  Nonetheless, this
neutrino flux is high enough to induce many convective overturns within
$\sim 100$ km of the neutron star (Thompson 2000, hereafter \cite{T00};
\S \ref{secfive}),
and thus to wind up a seed magnetic field within the accretion flow.
As the result of the 

\section{Amplification of Magnetic Fields}\label{sectwo}

The supernova core is subject to a violent convective instability
inside the $\nu$-sphere, driven by optically thick neutrino cooling;
and a distinct convective instability below the shock, triggered by
heating by the outward-streaming neutrinos (Fig. 1).  This second instability
will persist as long as substantial accretion continues onto the
neutron core, and the core remains a luminous source of neutrinos.  We
first show that entropy gradients will drive convection below the
neutrinosphere of a young neutron star. We then consider the response
of a magnetic field to instabilities either above or below the
neutrinosphere.

\subsection{Convection Inside the Neutrinosphere}\label{inside}

The stability of a neutron star to convection depends on gradients
in both entropy and composition.  The key compositional parameter
is the net lepton number per baryon, $Y_\ell = Y_e + Y_{\nu_e}$,
where $Y_e$ and $Y_{\nu_e}$ are respectively the numbers of electrons and
electron neutrinos per baryon.   During the first $\sim 10$ seconds,
when the star is optically thick to neutrinos, it is unstable to 
convection (\cite{WM};  Burrows 1987; TD93; \cite{KJM}; \cite{pons}).  
The Ledoux criterion for convective instability is expressed 
most conveniently as
\be\label{ledcon} 
{dS\over dR} + {(\partial\rho/\partial Y_l)_{P,S}\over 
(\partial\rho/\partial S)_{P,Y_l}}{dY_l\over dR} = 
{dS\over dR} - \left({\partial S\over\partial Y_l}\right)_{P,\rho}
 \,{dY_l\over dR} < 0.  
\ee 
(e.g. \cite{lm}).

It is worth explaining here why this deep convective instability is generic,
because its existence has been called into question on the basis
of hydrodynamical simulations of the outer neutron core, at relatively
low optical depth to neutrinos (\cite{Mezz98}).
Deep convection is distinct from the short-lived
overturn which occurs where the weakening shock establishes a negative
entropy gradient and which, according to some simulations such as
those of \cite{BHF} and Bruenn \& Mezzacappa (1994), may be quenched by 
neutrino diffusion in the first 20-50 ms after bounce.  In deep convection,
the gradients of $Y_l$ and $S$ both tend to be reduced (e.g. \cite{KJM}),
but are not entirely eliminated.  Rather, they are constantly regenerated by
radiative losses from the neutrinosphere.
Thus, convection must continue until the magnitude of
the radiative gradient drops below the adiabatic gradient. The
reduction of the radiative gradient results primarily from the decrease
in flux over the Kelvin time, which is of order ten seconds.

Ledoux convection is always encouraged by negative $dS/dR$, but the
effect of $dY_l/dR$ on convective instability depends on the temperature
and composition.  Negative $dY_l/dR$ is quickly established in 
a cooling neutron core by neutrino transport  at high $Y_l$  and by
chemical equilibrium between electrons, protons and neutrons at
at low $Y_l \la 0.1$ (where the lepton number is carried almost entirely by 
the electrons).  The sign of the thermodynamic derivative
$(\partial S/\partial Y_\ell)_{P,\rho}$ is therefore crucial to the nature
of the convective instability.   The negative lepton-fraction gradient
is de-stabilizing [$(\partial S/\partial Y_\ell)_{P,\rho} < 0$] if
the neutrino chemical potential $\mu_{\nu_e}$ is high enough 
{\it and} if the temperature is lower than a critical value
$T_\star$. 
In this situation, where a compositional gradient is potentially the
main driving force behind the convective instability,
salt-finger effects may play an important role in its non-linear
development (\cite{WM}).  

The critical temperature $T_\star$ can be estimated analytically
when the protons, neutrons, electrons, and electron neutrinos are all 
treated as ideal and almost degenerate fermi gases.  As derived
in Appendix A.2 (eq. [\ref{tstarb}]), one has\footnote{Neglecting 
terms of the order of 
$Y_{\nu_e}/Y_e$ and $\mu_e/m_nc^2$, but not 
$\mu_{\nu_e}/\mu_e = (2Y_{\nu_e}/Y_e)^{1/3}$.}
$$
k_BT_\star \simeq {\mu_e\over \pi}\,
\left[
1-\left({Y_e\over 1-Y_e}\right)^{2/3} 
\right]^{-1/2}\,
\left[{3(2\mu_{\nu_e}-\mu_e)\over m_nc^2}\right]^{1/2}
$$
\be\label{tstar} 
 = 35\,\left({n_b\over n_{\rm sat} }\right)^{1/2}\,
\left({2\mu_{\nu_e}-\mu_e \over \mu_e}\right)^{1/2}\;{\rm MeV} 
\;\;\;\;\;\;(Y_e \simeq 0.1).
\ee 
Here $\mu_e$ and $\mu_{\nu_e}$ are the chemical
potentials of the electrons and electron neutrinos, respectively.  We
have scaled the density to the nuclear saturation density
$n_{\rm sat} = 1.6\times 10^{38}$ cm$^{-3}$ (corresponding to
a mass density $m_n n_{\rm sat} = 2.8\times 10^{14}$ g cm$^{-3}$).  
This critical temperature is comparable to the initial
peak temperature in the neutron core.  Negative $dY_l/dR$ induces
convection below the temperature $T_\star$, but only as long as
$\mu_{\nu_e} > {1\over 2}\mu_e$, or equivalently
\be 
Y_{\nu_e} > {1\over 16} Y_e.
\ee 
In fact, $Y_{\nu_e}$ falls below this bound quickly in the outer
parts of the neutron core, in about 1 second according to the 
cooling models of Pons et al. (1999).  Thereafter, the 
composition gradient becomes stabilizing (\cite{lm}; \cite{gr};
\cite{KJM}; see \S \ref{brunt}).  
However, we now show that a negative entropy gradient can drive convection
in a multicomponent, noninteracting Fermi gas,
even in the presence of a stabilizing composition gradient.

Such a ``late phase'' of
convection occurs if the radiative temperature gradient
is sufficiently super-adiabatic. The radiative gradient is set by the
neutrino opacity, the energy flux, and the temperature:
\be \label{radiative} 
F_{\rm rad}=-\left({7N_\nu\over 8}\right)\,
{16\sigma_{\rm SB} T^4\over 3}\left(1\over \langle
n_b\sigma_\nu\rangle R\right) {d\ln T\over d\ln R}.
\ee 
Here, $\sigma_\nu$ is the neutrino cross section and the
angular brackets indicate a Rosseland mean.  During the intermediate
stages of cooling, the energy flux is carried primarily by 
$\mu$ and $\tau$ neutrinos:  their opacity is due to scattering
and is smaller than the absorption opacity of $\nu_e$ on neutrons
above a temperature of $\sim 10$ MeV (where the absorption
is not Fermi-blocked by the degenerate electrons; \cite{pons}).
In this situation, the effective number of neutrino species in
expression (\ref{radiative}) is $N_\nu \simeq 2$.  
However, the production rate of $\mu$ and $\tau$ neutrinos
may be so low that they cannot carry the flux, in which case $N_\nu=1$.

A negative entropy gradient is generated by steady radiative transport,
as is easily seen from the following argument.  The 
opacity of non-degenerate neutrinos scattering from degenerate nucleons 
scales as $E_\nu^3$ (as compared with the $E_\nu^2$ energy dependence
of the weak neutral current in a non-degenerate plasma; \cite{Iw81}).
The Rosseland mean free path is $\langle\sigma\rangle(T) \simeq
\sigma(E_\nu= 2k_BT)$ for a Fermi-Dirac distribution 
of neutrinos with temperature $T$ and vanishing chemical potential.  
The mean-free path for neutron scattering is
\be
{1\over n_n\sigma(E_\nu)} = 2\times 10^3\,\left({E_\nu\over 
30~{\rm MeV}}\right)^{-3}\;\;\;\;\;\;{\rm cm}
\ee
at a density $n_b = n_{\rm sat}$
(we normalize to Fig. 12 of \cite{reddy}).  The radiative temperature
gradient can, then, be directly related to the total radiative flux 
$F_{\rm rad} = L_\nu/4\pi R^2$:
\be
\left({d\ln T\over d\ln R}\right)_{\rm rad} = 
-1.3\,\left({N_\nu\over 2}\right)^{-1}\,
\left({L_\nu\over 10^{52}~{\rm erg~s^{-1}}}\right)\,
\left({T\over 10~{\rm MeV}}\right)^{-1}\,
\left({R\over 10~{\rm km}}\right)^{-1}.
\ee

This radiative gradient must be compared with the 
gradient required to trigger convection in the face of a 
(possibly) stablizing composition gradient.  This convective
temperature gradient is calculated by setting $dS/dR = 
(\partial S/\partial Y_l)_{P,\rho}dY_l/dR$.  The entropy (per baryon)
is provided mainly by free neutrons and protons, with respective
abundances $Y_n \simeq 1-Y_e$ and $Y_p = Y_e$:
\be
{S\over k_B} \simeq {\pi^2\over 2}\left[Y_e
\left({k_BT\over \mu_p}\right) + 
(1-Y_e)\left({k_BT\over \mu_n}\right)\right].
\ee
Here, the $\mu_p$, $\mu_n$ are the chemical potentials of the
degenerate protons 
and neutrons respectively.
This critical temperature gradient for convection is calculated
in Appendix A.3:
$$
\left({d\ln T\over d\ln R}\right)_{\rm conv} \simeq
{2\over3}\left({d\ln\rho\over d\ln R}\right) -
{Y_e^{1/3}\over
Y_e^{1/3}+(1-Y_e)^{1/3}}\times\;\;\;\;\;\;\;\;\;\;\;\;\;
$$
\be \label{convective} 
\times\left[
3Y_e\,{\hbar^3 c n_b\over m_n (kT)^2}\,
\left({\mu_e-2\mu_{\nu_e}\over\mu_e}\right)
+ {1\over 3}
\left\{
1 - \left({Y_e\over 1-Y_e}\right)^{2/3}
\right\}
\right]
{d\ln Y_e\over d\ln R},
\ee 
Notice that the second term (proportional to
$d\ln Y_e/d\ln R$) is guaranteed to be stabilizing when 
$d\ln Y_e/d\ln R < 0$ and $2\mu_{\nu_e} < \mu_e$. The coefficient
\be
3\,Y_e{\hbar^3 c n_b\over m_n (kT)^2} = 1.0\,
\left({n_b\over n_{\rm sat}}\right)\,\left({Y_e\over 0.1}\right)\,
\left({k_BT\over 20~{\rm MeV}}\right)^{-2}.
\ee

This analytical criterion indicates a slightly earlier end to
convection than is found in the cooling models of Pons et al. (1999)
(cf. their Figs. 9, 21, and 24).    
The value of $(d\ln T/d\ln r)_{\rm conv}$ depends on a cancellation
between two terms of opposing signs (on the RHS of  eq. [\ref{convective}]).
At a density $n_b \simeq n_{\rm sat}$
and a density gradient $d\ln\rho/d\ln r \simeq -3$, 
the radiative temperature gradient becomes too weak to induce
convection below a luminosity  $L_\nu \simeq 1\times 10^{52}$ erg s$^{-1}$
and temperature of $\sim 20$ MeV.  Since the binding energy is
$3\times 10^{53}$ ergs, the luminosity will be high enough to drive
convection for 10 seconds or more. 



Unlike a main sequence star, the thermal energy in the young neutron
star is not generated in the central core.  Heat is generated through
the entire star as the lepton fraction relaxes to its equilibrium value.
The absence of shock heating in the inner half of the star leads to
a positive temperature gradient at small radii, which delays the onset
of convection deep in the star (\cite{bl}). 
However, as the outer layers cool by neutrino loss, a
negative entropy gradient is produced at smaller and smaller radii.

The above argument shows that passive neutrino transport through a newly
formed neutron star leads inevitably to convective instability.  
Once the inconsistency of purely radiative heat transport is established,
one knows that some fraction of the heat flux must be carried by
convection.  In other words, the negative entropy gradient is not entirely 
erased by the convection (as is apparent in the simulations of KJM).
Rather, it is maintained at a large enough magnitude to carry the flux
needed to supply that lost at the neutrinosphere -- until the
magnitude of the radiative gradient drops due to decreasing luminosity
or increasing neutrino mean free path.

The convective energy flux is easily estimated using mixing length theory.
In terms of the convective velocity $V_{\rm con}$, one has
\be 
F_{\rm con} \simeq \rho V_{\rm con}^3.
\ee 
Numerically, $V_{\rm con}\approx10^8\,{\rm cm\,s^{-1}}$, which is
much less than than the speed of light and substantially smaller than
the sound speed $c_s\approx 5\times10^9\,{\rm cm\,s^{-1}}$.
Notice that the convective velocity varies as $F_{\rm con}^{1/3}$ and so
depends weakly on the convective efficiency.  As we review in the next
section, this means that {\it the maximum possible dynamo-generated magnetic
field will also vary weakly with the convective efficiency}.


The latest numerical simulations reach differing conclusions about the
persistence of Ledoux convection inside the $\nu$-sphere, but for
reasons that can be ascribed in part to the handling of neutrino
transport.  KJM employ a fully two-dimensional hydrodynamic code, with
neutrino transport restricted to radial rays, and find that convection
continues over the entire Kelvin time.  The simulations of
Mezzacappa et al. (1998)
further restrict neutrino transport to one spatial
dimension, and find that the two-dimensional convective motions die
out after a few tens of milliseconds.  They argue that Ledoux
convection is inhibited by neutrino transport in the outer layers of
the neutron core ($\rho \lsim 10^{13}$ g cm$^{-3}$), where the
diffusion time $\sim \tau_{\nu_e} \ell_P/c$ is shorter than the
convective overturn time in the absence of local neutrino heating and
cooling.  However, deeper in the core the convection overturn time
$\tau_{\rm con} = \ell_P/V_{\rm con} \sim 1-3$ ms is much shorter than the
Kelvin time and this conclusion is reversed.  The cooling
models of Pons et al. (1999), which include more realistic neutrino
opacities, find that a deep convective instability persists for tens
of seconds.  

A recent paper by Miralles, Pons, and Urpin (2000) carefully treats
how convective instability is modified by the transport of heat and lepton
number, and by viscosity.  They find that the cooling neutron star
simulated by Pons et al. (1999) is subject to a slower double-diffusive
instability (the `neutron finger instability'; \cite{WM}) even in regions
that are stable according to the Ledoux criterion (\ref{ledcon}).  
Nonetheless, at both 1 s
and 20 s, the region of the star that is subject to an unstable g-mode
with a fast growth rate ($\la 1$ msec; Fig. 3 of Miralles et al. 2000) 
closely approximates the region found previously 
to be unstable according to the Ledoux criterion
(Fig. 24 of \cite{pons}).  The new `convective'
instability criterion (eqs. [25] and [26] of Miralles et al. 2000) is
more restrictive than Ledoux, so that regions of the star that
are unstable according to eq. (\ref{ledcon}) are labelled as being
subject only to the `neutron finger' instability (in spite of the
equally fast growth rate).  In fact, one can show that this new
`convective' instability criterion remains more restrictive 
even in the limit of slow transport, where the Ledoux criterion should
be accurate.  For that reason, we interpret the results of
Miralles et al. (2000) to imply that transport effects {\it do not}
significantly restrict the region of a newborn neutron star that 
undergoes a vigorous convective instability.

\subsection{Dynamical Balance between Convective and Magnetic 
Stresses}\label{dynbal}

Even in the absence of rotation, these convective motions carry enough
energy to amplify a magnetic field to enormous strengths.  The argument
for such a stochastic dynamo is based on a {\it phenomenological scaling},
rather than purely theoretical considerations, and is presented in detail in
TD93.  One observes in the convective envelope of the Sun a small
scale (`intranetwork') component of the magnetic field, with a mean
pressure about 10 percent of the turbulent pressure, \be\label{beq}
{\langle B^2\rangle/4\pi\over \rho V_{\rm con}^2} \equiv \varepsilon_B
\sim 0.1.  \ee This field is present even at Solar minimum, when it is
distributed almost uniformly across the Solar disk (\cite{murray}).  It
has been conjectured to be the consequence of a stochastic,
non-helical dynamo that builds up magnetic flux on the scale of an
individual convective cell (TD93; \cite{DYR}).  Such a dynamo
would operate almost independently of the global Solar dynamo that
manifests itself at the surface in the form of sunspot activity and
the dipolar magnetic field\footnote{The two dynamos are not, of
course, entirely independent: the global dynamo feeds magnetic flux
into the convection zone.  However, one infers from the time-variation
of Solar p-mode frequencies with Solar cycle that the proportionality
between magnetic and turbulent stresses is maintained down to least
six pressure scaleheights below the photosphere (\cite{GMWK}).  This
strongly indicates a local amplification mechanism 
that leads to a dynamical balance between the two stresses.}.  Some
theoretical support for this conjecture comes from models of dynamo
action in mirror symmetric turbulence (\cite{RS}; \cite{MP}).

The total convective kinetic energy grows from $E_{\rm con} \sim 0.5$
to $1\times 10^{49}$ ergs between 0.1 and 1 s, as the
convective zone expands to encompass the entire neutron star
(KJM).  At 0.1-0.2 s after bounce, a zone interior
to $0.9 M_\odot$ is unstable to Ledoux convection, with overshoot 
extending
out to an enclosed mass of $1.05 M_\odot$. 
This is in accord with a simple mixing length argument (\cite{burrows87}; 
TD93).
The convective luminosity is related to the luminosity $L_\nu
= L_{\nu\,52}\times 10^{52}$  erg s$^{-1}$ in {\it all}
neutrino species via $L_{\rm con} \sim L_\nu \sim 4\pi R_{\rm con}^2 
\rho V_{\rm con}^3$.   The kinetic energy in convection  $E_{\rm con}  =
{4\pi\over 3} R_{\rm con}^3\times {1\over 2}\rho V_{\rm con}^2$
can then be expressed in terms of the convective overturn time
\be\tau_{\rm con} \equiv {\ell_P/V_{\rm con}} = 2\times 10^{-3}\,
\left({R_{\rm con}\over 30~{\rm km}}\right)\,
\left({V_{\rm con}\over 3\times 10^8~{\rm 
cm~s^{-1}}}\right)^{-1}\;\;\;\;\;\;
{\rm s}
\ee
via
\be 
E_{\rm con} = L_{\rm con} \tau_{\rm con}
\sim \Bigl({1\over 2}-1\Bigr)\times 10^{50}\, \left({L_{\nu\,52}\over 5}
\right)\,\left({\tau_{\rm con}\over 2~{\rm ms}}\right)\;\;\;\; {\rm ergs}.
\ee 
The pressure scale height $\ell_P \sim R/5$ while the nucleons are
only mildly degenerate.  Note that the convective overturn time is
short compared with the expected initial spin periods of most neutron
stars (\cite{dt92}).  Applying the scaling (\ref{beq}) for the
magnetic energy implies a total magnetic energy
\be 
E_B \equiv \int^{R_\nu} {B^2\over 8\pi}  d^3 x = \varepsilon_B E_{\rm con}
\sim 10^{49}\,\left({\varepsilon_B\over 0.1}\right)\,
\left({E_{\rm con}\over 10^{50}~{\rm ergs}}\right)\;\;\;\; {\rm ergs}
\ee 
inside the $\nu$-sphere.  The corresponding r.m.s. magnetic field is
\be\label{brms} 
\langle B^2\rangle^{1/2}_{R_\nu} = \left({6E_B\over R^3}\right)^{1/2}
\sim 1.5\times 10^{15}\,\left({\varepsilon_B\over 0.1}\right)^{1/2}\,
\left({E_{\rm con}\over 10^{50}~{\rm ergs}}\right)^{1/2}
\,\left({R_\nu\over 30~{\rm km}}\right)^{-3/2} \;\;\;\;{\rm G}.
\ee 

It should be noted that this dynamo process taps the energies of the
diffusing $\nu_\mu$ and $\nu_\tau$, as well as the $\nu_e$,
$\bar\nu_e$.  By contrast, direct neutrino heating outside the
$\nu$-sphere is mediated by absorption of $\nu_e$ and $\bar\nu_e$
only.

The region below the shock provides a second potential site for
magnetic field amplification.  Heating by the outward flux of
electron-type neutrinos, through the charged-current absorption
$\nu_e (\bar\nu_e) + n (p) \rightarrow e^- (e^+) + p (n)$,
induces strong convection (e.g. Burrows et al. 1995; \cite{janka96}).   
The conditions are especially favorable
for magnetic field amplification if prolonged infall continues after the 
shock succeeds (leading to the formation of a secondary accretion shock)
or if the expansion of the shock is asymmetric, so that accretion continues in
one hemisphere even as the shock expands in the opposing hemisphere
(\cite{T00}).  Buoyancy forces will pin
magnetized material just below the shock even while heaver material
settles below the shock toward the neutrinosphere.  The convective
Mach number is expressed in terms of the accretion rate $\dot M$,
shock radius $R_{\rm sh}$, core mass $M_{\rm core}$, and the (normalized)
electron neutrino luminosity ${\cal L} = Y_n L_{\bar\nu_e}\,
(\langle \varepsilon^2_{\bar\nu_e}\rangle/200~{\rm MeV}^2) + Y_p
L_{\nu_e}\,(\langle \varepsilon^2_{\nu_e}\rangle/200~{\rm MeV}^2)$,
\be\label{rhotwo} 
(Ma)_{\rm con} = 0.6\,\varepsilon_{\rm heat}^{1/3}\,
\left({{\cal R}\over 10}\right)^{1/2}\,
\left({R_{\rm sh}\over 100~{\rm km}}\right)^{-1/6} \,
\left({{\cal L}_{52}\over 4}\right)^{1/3}\,
\left({M_{\rm core}\over 1.4~M_\odot}\right)^{-1/2}.
\ee 
Here $\varepsilon_{\rm heat}$ is the dimensionless ratio of the
net heating rate (from which cooling by electron and positron captures 
has been subtracted) to the gross heating rate (uncompensated by 
neutrino cooling).  The  convective overturn time is 
$\tau_{\rm con} \simeq 2\times 10^{-3}\, \varepsilon_{\rm heat}^{-1/3}\,
(R_{\rm sh}/100~{\rm km})^{5/3}\,({\cal
L}_{52}/4)^{-1/3}$ s (\S 2.3 of \cite{T00}).  
This overturn is fast enough to allow many efoldings of
the buoyant magnetic field, if the shock radius $R_{\rm sh}$ has collapsed
to $\sim 100$ km.  When the magnetic stresses reach a dynamical
balance with the convective stresses, according to equation
(\ref{beq}), the strength of the field has increased to (\cite{T00})
\be\label{brmsb} 
\langle B^2\rangle^{1/2} = 1\times
10^{14}\,\varepsilon_{B\,-1}^{1/2}\,\varepsilon_{\rm heat}^{1/3}\, 
\left({\dot M\over M_\odot ~{\rm s^{-1}}}\right)^{1/2}
\left({R_{\rm sh}\over 100~{\rm km}}\right)^{-17/12}\,
\left({{\cal L}_{52}\over 4}\right)^{1/3}\,
\left({{\cal R}\over 10}\right)^{1/2}\, \left({M_{\rm core}\over
1.4~M_\odot}\right)^{3/4}.  
\ee 

\subsection{Convective Dynamo During Late Infall}

It is also worthwhile estimating how this equipartition field will
diminish, as the rate of accretion drops off.  Late infall
($\Delta M = O(10^{-1})\,M_\odot$ over $\sim 10^3-10^4$ s) 
is a plausible consequence of the formation
of a rarefaction wave in the exploding supernova core 
(\cite{ww95}; Fryer, Colgate, \& Pinto 1999).  Let us
suppose that $\dot M$ remains large enough that neutrino cooling at 
the base of the accretion flow continues to be dominated by electron/positron
captures on free protons and neutrons, as compared with $e^\pm$ annihilation 
$e^+ + e^- \rightarrow \nu + \bar\nu$ (as assumed in the models of 
Chevalier 1989).  We focus on the inner regions of the flow,
$R \sim 30-100$ km, well inside the accretion shock.  Here the 
temperature is high enough for helium to be photo-dissociated,
but the capture rates on free $n$, $p$ are still small enough to 
prevent the formation 
of a stabilizing gradient $dY_e/dR$.  The thermal pressure is then
dominated by photons and (at least mildly) relativistic $e^\pm$ pairs, 
$P = {11\over 12} aT^4$.  The corresponding 
isentropic hydrostatic profile is $T(R) \propto R^{-1}$,
$\rho(R) \propto R^{-3}$, and $P(R) \simeq {1\over 4} (GM_{\rm NS}/R) \rho(R)
\propto R^{-4}$.  The base of this settling flow may be defined as the 
radius where the flow time $({1\over 4}R/V_R)_R$ across a pressure 
scale-height equals the time $(3P/\dot Q^{eN})_R$ for $e^\pm$ captures 
to remove the thermal energy.  (The neutrino emissivity
per unit volume scales as  $\dot Q^{eN} \propto T^6\rho$ 
in a medium with non-degenerate electrons.)  Cooling is slower than
advection outside this radius, and the gradient $dY_e/dR$ is too 
weak to stabilize the flow against the effects of neutrino heating
(\cite{Mezz98}).  As a result the flow becomes violently convective.

Combining the constraint of mass conservation, $\dot M = 4\pi R^2 V_R \rho$,
with the equation of hydrostatic balance at the base of the settling flow, 
and taking the cooling radius to be a fixed multiple of the stellar
radius\footnote{In the presence of an extended atmosphere, the
stellar radius is conveniently defined as the radius where
$\rho = \rho_{\rm sat}$.}, one finds the scaling relations
\be\label{rhotscal}
\rho(R) \propto \dot M^{2/5}; \;\;\;\;\;\; T(R) \propto \dot M^{1/10}
\ee
at a fixed radius $R$.

From the relations (\ref{rhotscal}), one can deduce several things:

1.  The temperature at the base of the settling flow typically exceeds 
$m_ec^2$, for parameters of interest, as is needed for the pressure
to be dominated by photons and relativistic $e^\pm$ pairs.
For example, the temperature is $\sim 1$ MeV at the base of an 
accretion flow carrying a mass flux $\dot M \sim 10^{-4}\,M_\odot$ s$^{-1}$.

2.  The nucleons in the settling flow become less degenerate with
decreasing accretion rate:  the ratio of temperature to fermi energy
scales as $k_B T/\varepsilon_p \propto \dot M^{-1/6}$.
At the same time, the thermal energy per baryon remains almost constant
at fixed radius, $T^4/\rho \propto \dot M^0$, and so effective 
dissociation of helium is expected at the base of the accretion flow, 
even at relatively low accretion rates.

3.  At an intermediate infall rate $\sim 10^{-4}-10^{-5}\,M_\odot$ s$^{-1}$, 
the assumption of neutrino cooling by $e^\pm$ captures is indeed
self-consistent (as compared with $e^\pm$ annihilation 
$e^+ + e^- \rightarrow \nu + \bar\nu$).    The relative 
rates of energy loss through these two channels vary weakly with 
$\dot M$ at a fixed radius.  In particular, $\dot Q^{eN}/\dot Q^{e^\pm} 
\propto \rho/T^3 \propto \dot M^{1/10}$, before $e^\pm$ captures have 
removed enough heat to make the electrons degenerate.

4.  Neutrino heating still causes the infalling material to be
violently convective, before finally settling onto the neutron
star surface (\cite{T00}).  At late times, the luminosity in
electron-type neutrinos can be directly related to $\dot M$,
\be\label{lnulate}
L_{\nu_e} + L_{\bar\nu_e} =
{G\dot M M_{\rm NS}\over R_{NS}} =
7\times 10^{49}\,\left({\dot M\over 10^{-4}\,M_\odot {\rm s^{-1}}}\right)
\,\left({M_{\rm NS}\over 1.4\,M_\odot}\right)\,
\left({R_{\rm NS}\over 10~{\rm km}}\right)^{-1}
\;\;\;\;\;\;{\rm  erg s^{-1}}.
\ee
The heating rate (per unit volume) due to charged-current absorption
of these neutrinos on the settling baryons is $3P/\tau_{\rm heat} 
\propto \rho\dot M$ (eq. [\ref{theat}]).  The energy density in
the resulting convective motions scales as ${1\over 2}\rho V_{\rm con}^2
\simeq (3P/\tau_{\rm heat}) \times (\ell_P/V_{\rm con})$.  As a result, the 
convective speed decreases more slowly with decreasing accretion rate,
$V_{\rm con}(R) \propto \tau_{\rm heat}^{-1/3} \propto
\dot M^{1/3}$, than does the settling speed, $V_R(R) \propto \dot M^{3/5}$.
Moving outward through the accretion flow, the 
capture (cooling) timescale grows as $\sim \rho^{-1} T^{-2} \sim R^5$, given
an approximately isentropic hydrostatic profile $T \propto R^{-1}$
and $\rho \propto R^{-3}$.  By contrast, the heating time increases 
only as $\tau_{\rm heat} \sim R^2$.
Thus, on general grounds one expects a `gain region', within which
$dS/dR$ is strong enough, and $dY_e/dR$ is weak enough, to trigger
convection (e.g. Burrows et al. 1995; \cite{janka96}).

5. Finally, using the above scaling relations, it is easy to work
out how the equipartition magnetic field scales with accretion rate:
\be\label{eqbscale}
\langle B^2\rangle^{1/2} = \varepsilon_B^{1/2}\,(4\pi\rho)^{1/2} 
V_{\rm con} \propto \dot M^{8/15},
\ee
at a fixed radius.  The equipartition field drops by a factor 
$\sim 0.01$ from $\dot M = 1\,M_\odot$ s$^{-1}$ to 
$\dot M = 10^{-4}\,M_\odot$ s$^{-1}$.

\section{Transport of Magnetic Fields\label{transport}}

We now discuss the transport of a magnetic field through the
convective interior and convectively stable exterior of a nascent
neutron star.  The star's convective core is surrounded by an
overshoot layer of thickness $\Delta M \sim 0.1 M_\odot$ and another
convectively stable layer of similar mass that extends out to the gain
radius (KJM).  Magnetic fields amplified in the convective region are
transported rapidly into an adjoining overshoot layer (with a velocity
$\sim V_{\rm con}/\ell_P$).  Recent simulations of magetoconvection
(\cite{tobias}) show the the magnetic field will remain pinned at the
base of a convective layer, even against the action of buoyancy
forces.  Thus, we expect the overshoot layer above the convective core
of a neutron star to acquire magnetic fields as strong as those in the
core.  Transport of the magnetic field through the surrounding
convectively stable layer is greatly aided by i) the concentration of
the field into a dense fibril state and ii) rapid neutrino heating.  
We also discuss an additional subtlety involving turbulent pumping, 
driven by a radial gradient in the turbulent diffusivity.

\subsection{Turbulent Pumping within Convective Zones\label{pumping}}

A nascent neutron star and a late-type main sequence star both are
deeply convective.  In the outermost convective layers of a main
sequence star, the turbulent diffusivity {\it increases outward} due
to the recombination of hydrogen, which buffers the decrease of
temperature with decreasing density.  By contrast, there are no similar
ionization effects that enhance the neutrino scattering or absorption
cross sections in a neutron star and, as we now show, the turbulent 
diffusivity 
\be
\chi = {1\over 3}V_{\rm con}\ell_P
\ee
{\it decreases outward}.  Once the $\nu$-sphere shrinks to 
$\sim 15-20$ km, the pressure scale height with the surface layer of
mildly degenerate nucleons scales with density as $\ell_P = P/\rho g 
\propto \rho^{2/3}R^2/M(< R)$, where $M(<R)$ is the mass enclosed 
within radius $R$.   Constancy of the convective energy flux implies 
that $V_{\rm con} \propto \rho^{-1/3} R^{-2/3}$.  Combining these 
two expressions yields
\be 
\chi \propto {\rho^{1/3} R^{4/3}\over M(<R)}.
\ee 
One sees that $\chi$ decreases with radius in the outer part of the
convection zone, where $\rho$ decreases at almost constant $R$ and
$M(<R)$.  A similar conclusion holds for the convective shell in between the
gain radius and the shock.

A passive contaminant such as smoke, or a magnetic field, experiences
a net drift velocity in the direction of {\it decreasing} turbulent
diffusivity, $\vec V_{\rm drift} = -\vec\nabla(V_{\rm con}\ell_p)$, whose
magnitude is comparable to the convective velocity, $|V_{\rm drift}| \sim
V_{\rm con}$.  Magnetic flux ropes are convected {\it downward} in the
upper layers of the Solar convection zone, and their buoyant rise to
the surface is prevented, as long as $V_{\rm drift} \gsim
(a/\ell_p)^{1/2}\,B/\sqrt{4\pi\rho}$ (e.g.  Ruzmaikin \& Vainshtein
1978).  (Here $2a$ is the width of the flux rope.)  By
contrast, the mean drift velocity is directed {\it upward} at the top
of a neutron star convection zone, and the magnetic heating rate
should be correspondingly higher.  We emphasize that this effect is
more pronounced if the flux ropes are tied together into a larger
network, so that the stochastic motions of individual flux ropes
within individual convective cells are suppressed with respect to the
mean drift velocity.

A related effect, which has been studied in the context of Solar
magnetism, involves the asymmetry between cold downdrafts and hot
updrafts in a compressible, convective layer.  In 
the Solar convection zone the downdrafts are much
denser and narrower than the updrafts and, as a result, have
a higher vorticity.  Radiation-hydrodynamical simulations
(\cite{dorch00}) show that the magnetic field lines are wound
up in the downdrafts, and suggest a tendency for the field to be 
pumped toward the bottom of the convective layer.  The convective 
region of a newly formed neutron star is, however, supported by the
pressure of mildly degenerate neutrons, and is less compressible than
the Solar atmosphere.   Numerical simulations (KJM) suggest a greater
symmetry between downdrafts and updrafts than is inferred for the Sun.

\subsection{Transport Aided by a High Neutrino Flux\label{fivetwo}}

In this section, we examine how the buoyant transport of a magnetic
field through a supernova core is facilitated by the intense neutrino
flux out of the core.  The neutrinos have two principal effects, which
we examine in turn.  The first involves heating of the non-degenerate
material outside the $\nu$-sphere, which equilibrates the temperature
between a flux tube and its surroundings, and allows the tube
to rise buoyantly across a convectively stable layer.
The second effect involves transport
of a magnetic field through the degenerate neutron core after
convection has stopped and the core has become stably stratified.  In
such a situation, the transport rate is determined by the speed with
which the electron fraction returns to its beta-equilibrium value
(\cite{pethick}; \cite{gr}).  Long after the
neutron star has radiated most of its original heat, the rate of
modified URCA reactions may be kept high by magnetic dissipation
(Thompson and Duncan 1996).  During the first few seconds of the life
of a neutron star, direct URCA reactions are possible, and proceed at
an extremely high rate dominated by charged current absorption of
$\nu_e$ and $\bar\nu_e$ on free nucleons.

We now focus on the early stages of a supernova explosion, {\it
before} the success of the shock, and the onset of a low density,
neutrino-driven wind outside the $\nu$-sphere.  The calculation
separates into two cases, depending on whether the pressure is
dominated by nucleons or by radiation and relativistic pairs.
Curvature and tension forces, discussed in \S 4, are now ignored and
the rope is assumed to sit horizontally in a plane-parallel
atmosphere.

\subsubsection{A Nucleon Pressure-Dominated Zone Outside the 
Neutrinosphere}

The outer atmosphere of a newly formed neutron star is convectively
stable, with entropy and electron fraction both having positive radial
gradients, $dS/dR$, $dY_e/dR > 0$.  Free neutrons within this
layer absorb electron neutrinos at the rate
\be 
\dot E_{\nu_e + n\rightarrow p+e^-} = {390\over 
f}\,\left({L_{\nu_e\,52}\over
4}\right)\,\left({\langle\varepsilon_{\nu_e}^2\rangle\over 200~{\rm 
MeV}^2}
\right)\,\left({R\over 100~{\rm km}}\right)^{-2}\;\;\;\;\;{\rm MeV\,
neutron^{-1}\,s^{-1}}
\ee 
(\cite{bw}).  The dependence on the mean square neutrino energy is
derived from the weak interaction cross section.  The expression for
proton heating ($\bar\nu_e + p\rightarrow n+e^+$) is obtained by
replacing $L_{\nu_e}$ and $\langle\varepsilon_{\nu_e}^2\rangle$ with
$L_{\bar\nu_e}$ and $\langle\varepsilon_{\bar\nu_e}^2\rangle$.  Here
$f\simeq {1\over 4}$ for isotropic emission at the $\nu_e$-sphere,
increasing to $f = 1$ further out where the neutrinos free stream.
The net heating rate due to absorption of $\nu_e/\bar\nu_e$ on free
neutrons ($Y_n$ per nucleon) and protons ($Y_p$ per nucleon) is
\be 
\dot Q^{\nu N} = \Bigl(Y_n\dot E_{\nu_e+n\rightarrow p+e^-} +
Y_p\dot E_{\bar\nu_e+p\rightarrow n+e^-}\Bigr){\rho\over m_n},
\ee 
per unit volume.  Heating is compensated by neutrino cooling driven 
(primarily) by electron and positron captures, so that the net heating
rate becomes $\dot Q^{\nu N} + \dot Q^{eN} = 
\dot Q^{\nu N}[1-(T/T_{\rm eq})^6]$.  The equilibrium temperature is
\be\label{teq} 
k_BT_{\rm eq} \simeq 2.35\,f^{-1/6}\,
\left({{\cal L}_{52}\over 4(Y_n+Y_p)}\right)^{1/6}
\left({R\over 100~{\rm km}}\right)^{-1/3}\;\;\;\;\;\;{\rm MeV}.
\ee 
Here, we define the convenient dimensionless parameter
\be 
{\cal L}_{52} \equiv Y_n\,L_{\nu_e\,52}\,
\left({\langle\varepsilon_{\nu_e}^2\rangle\over 200~{\rm MeV}^2}\right)
+ Y_p\,L_{\bar\nu_e\,52}\,
\left({\langle\varepsilon_{\bar\nu_e}^2\rangle\over 200~{\rm 
MeV}^2}\right).
\ee 
The time for the temperature to relax to the equilibrium (\ref{teq})
is $\tau_{\rm heat} \sim ({3\over 2}\rho k_BT/m_n)/6 \dot Q^{\nu N}$, or
\be\label{theat} 
\tau_{\rm heat} 
\simeq 6\times 10^{-4}\,f\,\left({T_{\rm MeV}\over 4}\right)\,
\left({R\over 50~{\rm km}}\right)^2\,
\left({Y_n L_{\nu_e\,52}+Y_p L_{\bar \nu_e\,52}\over 4}\right)^{-1}\,
\left({\langle \varepsilon_{\nu_e}^2\rangle\over 200~{\rm 
MeV^2}}\right)^{-1}
\;\;\;\;\;\;{\rm s},
\ee 
since the heat capacity is dominated by free nucleons.  The factor of
${1\over 6}$ arises from the $T^6$ dependence of the $e^\pm$-capture 
cooling
rate.  The effectiveness of neutrino heating can be measured by comparing
$\tau_{\rm heat}$ with the dynamical time $\tau_{\rm dyn} = \ell_P/c_s \sim
(R/5)(5k_BT/3m_n)^{-1/2}$:
\be 
{\tau_{\rm heat}\over \tau_{\rm dyn}} =
1.6\,f\,T_4^{3/2}\,\left({R\over 50~{\rm km}}\right)\,
\left({Y_n L_{\nu_e\,52}+Y_p L_{\bar \nu_e\,52}\over 4}\right)^{-1}\,
\left({\langle \varepsilon_{\nu_e}^2\rangle\over 200~{\rm 
MeV^2}}\right)^{-1}.
\ee 
This implies $\tau_{\rm heat} \lsim \tau_{\rm dyn}$ at the $\nu$-sphere
($R_\nu \sim 50$ km at 0.1 ms, decreasing to $R_\nu \sim 20$ km
at 1 s; e.g. \cite{MW}).

The nucleon pressure is suppressed inside a magnetic flux rope
that is in pressure equilibrium with its surroundings (Fig. 2.),
\be\label{peq} 
{B^2\over 8\pi} + {\rho_M k_BT_M\over m_n} = {\rho k_BT\over m_n}.
\ee 
However, the pressure of the electrons is only weakly perturbed
by the magnetic field when $Y_e$ is small.
The density is high enough that the electrons are
moderately degenerate ($\mu_e/T \gsim \pi$),
\be 
\rho \gg {(k_BT)^3 m_n\over (\hbar c)^3Y_p} =
2\times 10^8\,Y_p^{-1}\,T_{\rm MeV}^3\;\;\;\;\;\;{\rm g~cm^{-3}}.
\ee 
At a fixed radius, the equilibrium electron pressure is the result of a
competition between absorption of $\nu_e/\bar\nu_e$,
\be 
{\partial n_e\over\partial t} \propto  R^{-2}
\left[L_{\nu_e}\langle \varepsilon_{\nu_e}\rangle n_n -
L_{\bar\nu_e}\langle \varepsilon_{\bar\nu_e}\rangle n_p\right]
\simeq R^{-2} L_{\nu_e}\langle \varepsilon_{\nu_e}\rangle {\rho\over 
m_n}
\left(1-2Y_e\right),
\ee 
and the capture of (degenerate) electrons on nucleons,
\be 
{\partial n_e\over\partial t} \propto -\mu_e^6\rho \propto -Y_e^2\rho^3.
\ee 
The electron fraction therefore varies with density as
\be 
{1-2Y_e\over Y_e^2} \propto \rho^2,
\ee 
at fixed $R$.  The decrease $\Delta P_e$
in electron pressure accompanying a decrease $\Delta\rho$ in density is
small where $Y_e \ll 1$;
\be 
{\Delta P_e\over P_e} = {4Y_e\over 3(1-Y_e)}\,{\Delta\rho\over\rho}.
\ee 
Furthermore, $P_e$ is only a fraction of the nucleon pressure because
$Y_e$ is small where $\rho$ is large:
\be 
{P_e\over P_n+P_p} = {Y_e \mu_e\over 4k_B T} = 0.3\, 
\left({\rho\over 10^{12}~{\rm g~cm^{-3}}}\right)^{1/3}
\,\left({T_{\rm MeV}\over 4}\right)\,\left({Y_e\over 0.1}\right)^{4/3}.
\ee 

We conclude that the flux rope is less dense than its surroundings,
$\rho_M < \rho$, when its temperature rapidly equilibrates with its
surroundings.  Rapid neutrino heating, $\tau_{\rm heat} \lsim \tau_{\rm dyn}$,
brings the temperature inside the flux rope close to the ambient
value, $T_M \simeq T$.  The density deficit inside the rope is then
$\rho_M -\rho \simeq -\rho\beta_P$, where
\be
\beta_P = {B^2\over 8\pi nT}.
\ee
Balancing the buoyancy force
(integrated over the cross-sectional area $\pi a^2$ of the rope)
against the drag force $C_d\times 2a V_R^2$, yields the equilibrium
vertical speed
\be\label{ualf} 
V_R=\left({\pi\over 4 C_d}{a\over\ell_P}\right)^{1/2}\,
{B\over \sqrt{4\pi\rho}}.
\ee 
(cf. Parker 1979).

The vertical speed $V_R$ is determined by the rate of neutrino
heating when $\tau_{\rm heat} \gsim \tau_{\rm dyn}$.  The force balance
$\rho_M\,V_R^2/a \sim g(\rho -\rho_M)$ implies that the
density is almost uniform, $\rho_M \simeq \rho$, and that there
is a temperature deficit
\be\label{tdiff} 
T_M - T = - \beta_P\,T
\ee 
inside the rope.  As the rope moves upward, $T_M$ changes due to
neutrino heating (eq. [\ref{theat}]) and adiabatic cooling,
\be \label{tderiv} 
{1\over T_M}{dT_M\over dt} =
-{1\over \tau_{\rm heat}}{T_M-T_{\rm eq}\over T_M} + (\gamma-1){1\over \rho_M}
{d\rho_M\over dt}.
\ee 
Note that it is the equilibrium temperature (\ref{teq}), rather than
the ambient temperature $T$, that enters on the right side of
(\ref{tderiv}).  Combining the conditions of pressure equilibrium and
flux conservation, $B/\rho_M =$ constant, yields the following
relation between time derivatives,
\be\label{tderivb} 
{1\over T_M}{dT_M\over dt} = {1\over P}{d P\over dt} -
\left(1+2\beta_P\right){1\over\rho_M}{d\rho_M\over dt}.
\ee 
Equations (\ref{tdiff}), (\ref{tderiv}) and (\ref{tderivb}) together
yield
\be 
{1\over \tau_{\rm heat}}\,{\beta_P T_{\rm eq} -(1-\beta_P)(T-T_{\rm eq})\over T_M} 
=
{1\over P}{d P\over dt} - \left(\gamma+2\beta_P\right)
{1\over\rho_M}{d\rho_M\over dt} =
(\gamma-1){dS\over dt} -
{2\beta_P\over\rho_M}{d\rho_M\over dt},
\ee 
where $S$ is the entropy per baryon in units of $k_B$. 
The vertical rise time $\ell_P/V_R$ of the rope is obtained by
substituting $d/dt \rightarrow V_R(d/dR)$ and $\rho_M \simeq \rho$,
\be  
{\ell_P\over V_R} = \tau_{\rm heat}
{(1-\beta_P)T\over\beta_PT_{\rm eq}-(1-\beta_P)(T-T_{\rm eq})}\,
\left({\ell_P\over R}\right)\,\left[(\gamma-1)\,{dS\over d\ln R} -
2\beta_P\,{d\ln\rho\over d\ln R}\right].
\ee 
In the marginally failed shock model of Janka \& Mueller (1996) 
one finds $dS/d\ln R \simeq +12$, $d\ln\rho/d\ln R \simeq -4$, 
and $\ell_P/R \simeq {1\over 5}$ within the stably stratified
layer outside the $\nu$-sphere at 100 ms after bounce.  Thus,
\be\label{uneu} 
 {\ell_P\over V_R} \simeq
1.6\,\tau_{\rm heat}\,\left({1-\beta_P^2\over\beta_P}\right)
\ee 
when $T \simeq T_{\rm eq}$.

It is straightforward to see that the upward, buoyant rise of a closed
loop of magnetic flux will continue through the convectively stable
layer outside the $\nu$-sphere and reach the convective shell interior
to the shock.  We consider both cases $\beta_P \ll 1$ and $\beta_P
\simeq 1$ in turn.  In the first case, the radial transport time is
limited by neutrino heating (eq.  [\ref{uneu}]) and scales as
$\beta_P^{-1}f T R^2 \propto \beta_P^{-1}f R$.  The radial dependence
of $\beta_P$ is obtained by noting that the density inside the loop
scales as $\rho_M \sim \rho$, and the flux density as $B \sim
\rho_M^{2/3}$.  Thus, $\beta_P = B^2/8\pi P \propto \rho^{4/3}/P
\propto R^{-1/3}$ and (using the same scalings $\rho \propto R^{-4}$
and $P \propto R^{-5}$) the transport time increases with radius.
At the inner boundary of the convective shell ($R \sim 100$ km) the
transport time is only $\sim 1\times 10^{-3}\beta_P^{-1}\,
(L_{\nu_e\,52}/4)^{-1}\, (\langle
\varepsilon_{\nu_e}^2\rangle/200~{\rm MeV^2})^{-1}$ s.

Transport is even faster when $\beta_P \simeq 1$, since
buoyancy is effective even in the absence of neutrino heating.
Balancing $B^2/8\pi = P$, and making use of the vertical velocity
(\ref{ualf}) of a horizontal flux rope,  one finds
\be 
{\ell_P\over V_R} \propto {R^{3/2}\rho^{1/2}\over P^{3/8}}.
\ee 
This yields  $\ell_P/V_R \propto R^{11/8}$, using the same scalings for
$P$ and $\rho$.

\subsubsection{A Radiation Pressure-Dominated Zone Outside the 
Neutrinosphere}

We now consider transport of a magnetic field through a convectively
stable atmosphere whose pressure is dominated by relativistic pairs
and photons.  The zone just interior to the gain radius fits this
description.  The temperature of the settling flow exceeds the
equilibrium value $T_{\rm eq}$ (eq. [\ref{teq}]), and neutrino cooling
dominates heating.  The pressure equilibrium condition becomes
\be 
{B^2\over 8\pi} +{11\over 12}aT_M^4 \simeq {11\over 12}aT^4.
\ee 

In this situation, the nucleon density inside the flux rope is an
independent variable, which determines the strength of the buoyancy
force and the rate of neutrino heating.  When $T_M^4 \ll T^4$, the
equilibrium magnetic flux density decreases with radius as $B \propto
T^2 \propto R^{-2}$.  The density inside a rising, horizontal flux
rope decreases according to $\rho_M \propto B \propto R^{-2}$.  By
contrast, the density of the surrounding settling flow decreases much
more rapidly --- faster than $\rho(R) \propto R^{-3}$ because the
relativistic entropy per baryon ($S \propto T^3/\rho$) increases with
radius.  Thus, a flux rope will traverse several pressure scale
heights only if it starts off at a density well below that of the
ambient settling flow.  In such a situation, direct neutrino heating
is not required to ensure buoyancy.  In practice, an effective barrier
to magnetic buoyancy is not expected during the first $\sim 1$ s, 
because the flow remains radiation-pressure dominated only within a 
short distance below the gain radius.

Radiation pressure also has a significant effect on the equilibrium
shape of a magnetic arcade with fixed endpoints.
An arcade of magnetic flux, whose footpoints are separated by
an angle $\Delta\phi$, will open to infinity when $\Delta\phi$
exceeds the critical value (\ref{phic}) given below.   This critical angle
depends on the dimensionless parameter $\widetilde\ell_P
\propto \rho T/P$ (eq. [\ref{newellp}]), in such a way that
the critical footpoint separation decreases as photons and pairs
contribute more to the pressure.

\subsection{Transport in a Convectively Stable Neutron 
Core}\label{latet}

We now consider the transport of a magnetic field in the central
neutron core, after the core has become convectively stable.  The
electrically charged particles (protons and electrons) are tightly
coupled to the magnetic field on short timescales.  Transport of these
charged particles through the degenerate neutron fluid proceeds
extremely slowly, and at high temperatures is limited by
neutron-proton drag (Goldreich and Reisenegger 1992).  As a result,
the interior of a newborn neutron star is very nearly an ideal
magnetofluid.  For this reason, we consider the bulk hydrodynamical
motion of an isolated rope of magnetic flux.  As in the previous
section, we neglect curvature and tension forces and assume plane
parallel symmetry.

The neutron core becomes convectively stable as it cools because
$\beta$-equilibrium value of $Y_e$ has a negative gradient, and the
density of cool degenerate matter decreases with decreasing $Y_e$
(\cite{rg}; \S \ref{brunt}).  This happens, at the latest, when the
temperature drops sufficiently that neutrinos begin to escape directly
from the core.  A flux rope that starts out in $\beta$-equilibrium is
less dense than the surrounding fluid; as it rises buoyantly, it falls
out of $\beta$-equilibrium and its density equilibrates.  We treat the
neutrons, protons and electrons as three ideal, degenerate Fermi
fluids.  The departure from $\beta$-equilibrium is measured by the
chemical potential imbalance
\be 
\Delta\mu = \mu_p^M + \mu_e^M - \mu_n^M.
\ee 
In pressure equilibrium,
\be 
P_e^M + P_p^M + P_n^M + {B^2\over 8\pi}  = P_e + P_p + P_n,
\ee 
which can be rewritten as
\be\label{relone}
\Delta\mu_e n_e + \Delta\mu_p n_p + \Delta\mu_n n_n + {B^2\over 8\pi} 
= 0
\ee 
when the magnetic pressure is a small fraction of the total pressure.
(As before, the superscript $^M$ denotes the interior of the flux rope,
and $\Delta\mu_e = \mu_e^M-\mu_e$, etc.)
The condition of charge neutrality, $n_e^M - n_p^M = n_e - n_p = 0$,
leads to the relation
\be\label{reltwo}
2 {n_e\Delta\mu_e\over\mu_e} = {n_p\Delta\mu_p\over\mu_p},
\ee 
and the condition of neutral buoyancy, $n^M_n + n^M_p = n_n + n_p$,
leads to
\be\label{relthree} 
{n_n\Delta\mu_n\over\mu_n} +{n_p\Delta\mu_p\over\mu_p} = 0.
\ee 
One solves for $\Delta\mu$ by combining these three equations,
\be\label{delmu} 
\Delta \mu = \left[{2\mu_n\over (1-Y_e)\mu_e} - 1\right]\,
{B^2\over 8\pi n_e}.
\ee 
This expression can be simplified further after observing that
the shift in the charged particle density is related to $\Delta\mu$
via
\be 
\Delta\mu = {\mu_e\over 3n_e}\,
\left[{2\mu_n\over (1-Y_e)\mu_e} - 1\right]\,(n_p^M-n_p),
\ee 
so that
\be\label{yediff} 
{n_p^M-n_p\over n_p} =
{\Delta Y_e\over Y_e} = {3\over 4} {B^2\over 8\pi P_e}.
\ee 
This shift in the electron fraction is similar, in order
of magnitude, to that induced by the Lorentz force acting
on a homogeneous $n-p-e$ plasma without gravity
(Goldreich and Reisenegger 1992).

The vertical speed $V_R$ of the flux tube is limited by the rate at
which the charged particle density is driven to its
$\beta$-equilibrium value by absorption of $\nu_e$ and $\bar\nu_e$ on
free nucleons,
\be  
{\partial Y_e\over\partial t} + V_R {\partial Y_e\over\partial R}
  =  {\Gamma(\nu_e + n \rightarrow e^- + p)-
         \Gamma(\bar\nu_e + p \rightarrow e^+ + n)\over \rho/m_n}.
\ee 
Here $\Gamma$ is the rate per unit volume, so that
\be 
\Gamma(\nu_e + n \rightarrow e^- + p)-
         \Gamma(\bar\nu_e + p \rightarrow e^+ + n)
= \langle \sigma_{\nu_e n}\rangle n_{\nu_e} n_n^M c -
  \langle \sigma_{\bar\nu_e p}\rangle n_{\bar\nu_e} n_p^M c.
\ee 
In equilibrium,
\be 
{\langle \sigma_{\nu_e n}\rangle n_{\nu_e}\over
\langle \sigma_{\bar\nu_e p}\rangle n_{\bar\nu_e}}
 = {n_p\over n_n} = {Y_e\over 1-Y_e},
\ee 
and so
\begin{eqnarray}\label{gamdiff} 
\Gamma(\nu_e + n \rightarrow e^- + p)-
         \Gamma(\bar\nu_e + p \rightarrow e^+ + n)
&= \langle \sigma_{\nu_e n}\rangle n_{\nu_e}\,(n_n^M-n_n)c -
 \langle \sigma_{\bar\nu_e p}\rangle n_{\bar\nu_e}\,(n_p^M-n_p)c\nonumber\\
&= Y_e^{-1}\,\langle \sigma_{\nu_e n}\rangle n_{\nu_e}c\,
(n_n^M-n_n).
\end{eqnarray} 
Since $\Delta Y_e/Y_e$ is assumed to be small, one can express $V_R$ in 
terms
of the gradient of the unperturbed electron fraction,
\be\label{unew} 
V_R \simeq {\Gamma(\nu_e + n \rightarrow e^- + p)-
     \Gamma(\bar\nu_e + p \rightarrow e^+ + n)\over (dY_e/dR)\rho/m_n}.
\ee 
Combining equations (\ref{yediff}), (\ref{gamdiff}) with (\ref{unew}),
one obtains
\be\label{umax0} 
{V_R\over c} = {3\over 4Y_e}\,\left|{d\ln Y_e\over d\ln R}\right|^{-1}\,
\Bigl(\langle \sigma_{\nu_e n}\rangle n_{\nu_e} R\Bigr)\,
{B^2\over 8\pi P_e}.
\ee 
Note that $\langle \sigma_{\nu_e n}\rangle n_{\nu_e} R =
(n_{\nu_e}/n_n)\tau_{\nu_e} = 7\times 10^{-4}
\tau_{\nu_e}\,(T_{\rm MeV}/20)^3\,(n_n/n_{\rm sat})^{-1}$.
In this expression, $\tau_{\nu_e} = n_n\langle\sigma_{\nu_e n}\rangle R
\simeq 10^4\,(k_BT/{\rm 20~MeV})^3$ is the absorption depth 
at 15 MeV $\la k_BT \la$ 40 MeV (cf. Fig. 7 of Reddy et al. 1998).
Substituting these expressions into (\ref{umax0}), one obtains
\be\label{umax} 
{V_R\over c} \simeq 4\times 10^{-9}\,\tau_{\nu_e}\,
\left({B\over 1\times 10^{14}~{\rm G}}\right)^2\,
\left({T_{\rm MeV}\over 20}\right)^3\,\left({Y_e\over 0.05}\right)^{-7/3}\,
\left({n_n\over n_{\rm sat}}\right)^{-7/3}.
\ee 
At the Kelvin time of $\sim 3$ s, the density is $n_n \sim n_{\rm sat}$
and the temperature is $k_BT \sim 20$ MeV at an enclosed mass of 
$\sim 1\,M_\odot$ (Fig. 9 of \cite{pons}).
A flux rope is able to overcome the stable stratification from this
depth only if $V_R/c \ga {1\over 3}R_{\rm ns}/ct \sim 3\times 10^{-6}$,
where $R_{\rm ns} \sim 10$ km is the stellar radius. 
The {\it lower} bound on the flux density which can rise buoyantly is then
\be\label{bmin} 
B \gsim 3\times 10^{13}\,
\left({T_{\rm MeV}\over 20}\right)^{-3}\,\left({t\over 3~{\rm s}}\right)^{-1/2}
\;\;\;\;\;\;{\rm G}.
\ee 
Notice the strong dependence on $T$.  The direct URCA reactions
freeze out at temperatures below $k_BT \sim 10$ MeV, and so 
transport of the magnetic field to the stellar surface becomes 
ineffective at $t \ga 10$ s.

The net effect of this transport process is to exchange a region of
high flux density with a region of lower flux density that lies above
it.  Over such short timescales, the neutron matter moves with the
magnetic field.  A field that uniformly threads the neutron core does
not induce buoyancy forces and can exceed (\ref{bmin}).
Perhaps coincidentally, this bound is comparable to the dipole magnetic field
inferred for the Soft Gamma Repeater sources in the magnetar model
(Thompson and Duncan 1996; \cite{kouveliotou99}).

\section{Magnetostatic Equilibria}\label{secfour}

The convective core of a neutron star is surrounded by a convectively
stable layer that extends from large to small neutrino optical depth,
and straddles the neutrinospheres (Fig. 1).  A rope of magnetic flux that is
forced from below by convective overshoot into this layer can continue
to rise buoyantly above the neutrinosphere, through the effects of 
neutrino heating as discussed in \S \ref{transport}.  We now consider the 
equilibrium shape of such a rope when its two endpoints are pinned from 
below (Fig. 3).  The picture of the
magnetic field that we adopt is one of spatially intermittent fibrils
that are confined near the boundaries of convective cells below the
$\nu$-sphere.  Such a distribution is observed at the Solar
photosphere (\cite{stenflo}) and is inferred in the upper convection
zone from helioseismology (\cite{GMWK}).  It is also observed in
simulations of a turbulent, conducting fluid at high magnetic Reynolds
number (\cite{thelen00}).

Aside from its dramatically higher temperature and stronger gravity, the 
exterior of newly formed neutron star is distinguished
from the non-thermal atmosphere of the Sun in an important respect.
Before the shock succeeds, the pressure remains high enough
that magnetic flux originating inside the star will be confined to
narrow fibrils well {\it outside} the $\nu$-sphere.  To see this, 
we compare the pressure in the r.m.s. magnetic field (\ref{brms}) inside the
$\nu$-sphere, with the pressure of free nucleons near the $\nu$-sphere,
\be\label{prel} 
\langle{\beta}\rangle =
{\langle B^2\rangle_{R_\nu}/8\pi\over nk_B T}  =
.005\,\left({\varepsilon_B\over 0.1}\right)\,\left({\rho\over
10^{12}~{\rm g~cm^{-3}}}\right)^{-1}\,\left({T_{\rm MeV}\over 4}\right)^{-1}
\,\left({R_\nu\over 50~{\rm km}}\right)^{-3}.
\ee 
This low volume average of $\beta$ originates in the relatively
low convective Mach number of the supernova core (as compared with
the upper convection zone of the Sun), combined with the assumption of a
common dynamo efficiency $B^2/4\pi\rho V_{\rm con}^2$.

The field that pierces the $\nu$-sphere is not force free.   We now 
calculate the equilibrium shape of a flux rope that is underdense 
compared with its surroundings and confined by the external pressure, 
so that $B^2/8\pi\simeq P$.   The diameter of such a confined bundle of 
flux (of cross-sectional area $\pi a^2$) increases slowly with distance
$R$ from the center of the neutron core,
\be 
{a\over R} \propto R^{\beta/4-1} \sim  R^0-R^{1/4},
\ee 
since the pressure stratification outside the $\nu$-sphere is
$P(R) \propto R^{-\beta}$ ($\beta \sim 4-5$).

As a result, the pressure scale height is a much larger fraction
of the radius, $\ell_P/R \sim {1\over 5}$, near the $\nu$-sphere
of a nascent neutron star, than it is near the Sun's photosphere,
where $\ell_P/R_\odot \sim 10^{-4}$.  As a result, a magnetic flux rope 
is subject to a strong buoyancy force on {\it both} sides of the $\nu$-sphere.
To calculate the equilibrium 
shape of the rope, we first consider the case of a plane-parallel 
atmosphere.  A flux rope sitting in the $x-z$ plane (Fig. 3),  with 
tangent vector
\be\label{tang} 
\hat{\bf l} =  \cos\theta\hat{\bf x} + \sin\theta \hat{\bf z},
\ee 
is subject to a buoyancy force density
\be\label{fbuoy} 
(\rho_M-\rho){\bf g} \;\;= \;\; \beta_P\rho|g|\,\hat{\bf r},
\ee 
where
\be 
\beta_P \equiv {B^2\over 8\pi P}
\ee 
and $\rho_M$ is the density inside the flux rope.
In equilibrium, the normal component of (\ref{fbuoy}),
\be\label{fbuoyb} 
(\rho_M-\rho)\left[{\bf g}-({\bf g}\cdot\hat{\bf l})\hat{\bf 
l}\right],
\ee 
is balanced against the (normal) tension force density
\be  
{B^2\over 4\pi}{\partial\hat{\bf l}\over \partial l}.
\ee 
When  neutrino heating equilibrates the temperatures
inside and outside the flux rope (\S \ref{fivetwo}),
the trajectory of the rope obeys the simple equation,
\be 
{d\theta\over dl} = - {\rho g\over 2P_m}\cos\theta
= -{\cos\theta\over 2\ell_P}\,\left({P\over P_m}\right).
\ee 
Here $P$ is the total pressure, and $P_m = \rho T/m_n$ is the pressure
in free nucleons (assuming that $T > 1$ MeV and alpha particles are
absent).  A related expression has been written down by Parker (1979)
for a plane-parallel atmosphere with negligible pressure in
relativistic particles.  The main consequence of the pairs and photons
is to reduce the scale height from $\ell_P = P/\rho g$ to
\be\label{newellp} 
\widetilde\ell_P = {P_m\over\rho g} = \left({P_m\over 
P}\right)\ell_P.
\ee 
The separation between the two footpoints is
\be\label{delxfoot} 
\Delta x = \int \cos\theta dl = 2\int_{-\theta_0}^{\theta_0}
\widetilde\ell_P[l(\theta)] d\theta,
\ee 
and the maximum height attained by the arcade is
\be 
z_{\rm max} = 2\int_0^{\theta_0} \tan\theta 
\widetilde\ell_P[l(\theta)] d\theta,
\ee 
where $\theta_0$ is the inclination of the flux tube from the horizontal at
$z = 0$.  In an atmosphere with constant pressure scale height 
$\widetilde\ell_P$, these expressions simplify to
\be 
\Delta x = 2\widetilde\ell_P \theta_0 \leq 2\pi \widetilde\ell_P,
\ee 
and
\be 
z_{\rm max} = 2\widetilde\ell_P \ln\left({1\over\cos\theta_0}\right).
\ee 
The flux arcade expands to infinity as the footpoint separation
approaches the critical value 
\be
\Delta x_c = 2\pi\widetilde\ell_P,
\ee
at which the flux segments become vertical, $\theta_0 = {\pi\over 2}$.  A
similar effect will occur in any stably stratified, plane parallel
atmosphere.

Notice that radiation pressure can have a significant effect on the 
equilibrium shape of a magnetic arcade, since $\widetilde\ell_P
\propto \rho T/P$.  As a result, the critical footpoint separation 
decreases as photons and pairs contribute more to the pressure.

These results are easily generalized to an atmosphere with spherical
symmetry.  The shape of a flux arcade that sits in the equatorial
plane (with azimuthal angle $\phi$) is defined by the equation
\be 
{d\over dl}\left(\theta -\phi\right) = -{\cos\theta\over 
2\widetilde\ell_P},
\ee 
which simplifies to
\be 
{d\theta\over d\phi} = -\left({R\over 2\widetilde\ell_P}-1\right).
\ee 
In a spherical atmosphere with $\widetilde\ell_P/R$ a constant, the
flux arcade attains a maximum radius
\be 
R_{\rm max} = R_\nu
\left({1\over 
\cos\theta_0}\right)^{\left(R/2\widetilde\ell_P-1\right)^{-1}},
\ee 
and opens to infinity when the angular separation between the footpoints
approaches the critical value
\be\label{phic} 
\Delta\phi_c = {2\theta_0\over R/2\widetilde\ell_P -1} =
{\pi\over R/2\widetilde\ell_P -1}.
\ee 
For example, if the pressure is dominated by relativistic particles
then $\ell_P \simeq R/4$ and $\widetilde\ell_P$ is determined by the
post-shock entropy.  Substituting $\widetilde\ell_P \simeq R/5$ 
outside the $\nu$-sphere (around 100 msec after bounce), one deduces
a critical separation 
\be
\Delta\phi_c \simeq {2\pi\over 3}.
\ee

Magnetic arcades that extend beyond the $\nu$-sphere are not fixed
in position:  their footpoints can be expected to wander in response
to the convective motions below.  Since these footpoint motions occur on
the short timecale $\sim \tau_{\rm con}$, it is important to check whether
an individual arcade is able to reach a new equilibrium configuration
over the timescale $\sim \tau_{\rm con}$.  Otherwise, the magnetostatic 
approximation is not justified. 
The motion of the upper part of an arcade is driven by the tension
force $f_{\rm tension} = (\pi a^2)(B^2/8\pi R_c)$ (where
$R_c$ is the curvature radius and $\pi a^2$ the cross-section
of the flux rope) and is resisted by the hydrodynamical drag
force $f_{\rm drag} = 2C_d a \rho V^2$.   Given a footpoint separation
$\sim 2R_c$, the velocity of the rope (unimpeded by drag) is
$V \sim 2R_c/\tau_{\rm con}$, and the ratio of drag to tension forces is
\be 
{ f_{\rm drag}\over f_{\rm tension} } \;=\;
 0.6\,{1\over \beta_P}\,
\left({R\over 50~{\rm km}}\right)^2
\left({T_{\rm MeV}\over 4}\right)^{-1}\,
\left({2R_c\over\ell_P}\right)^2\,
\left({\tau_{\rm con}\over 3~{\rm ms}}\right)^{-2}\,
\left({a\over 0.5 R_c}\right)^{-1}.
\ee 
One sees that smaller flux arcades ($2R_c \sim \ell_P$) that sit near
the $\nu$-sphere ($R_\nu \sim 30-50$ km) will be able to follow the
footpoint motions, whereas larger arcades that extend to a  radius
$R \gg R_\nu$ will lag behind, and probably become very tangled. 

Self-reconnection is also possible for smaller flux arcades that
expand into the convective layer that sits below the shock;
or for those which rotate rigidly with the central
neutron star and experience a ram pressure force density
\be 
\rho(R) (\Omega_{\rm core} R)^2
\ee 
that exceeds the buoyancy force density (\ref{fbuoy}).
This second condition is satisfied beyond a radius
\be 
R_{\rm max} = 320\,\left({P_{\rm core}\over 100~{\rm ms}}\right)^{2/3}\,
\left({M_{\rm core}\over M_\odot}\right)^{1/3}\,
\left({\rho-\rho_M\over\rho}\right)^{1/3}\;\;\;\;\;\;{\rm km},
\ee 
where $P_{\rm core} = 2\pi/\Omega_{\rm core}$ is the rotation period
of the neutron core.
However, as discussed in Chevalier (1989) and 
Thompson (2000), the neutron star may acquire
its rotational  angular momentum only during the last stages of
accretion from the pre-supernova core.

\section{Implications for Pulsars}\label{secfive}

This paper has investigated the response of a magnetic field to
the violent fluid motions and intense neutrino flux within a supernova core.
We can summarize our conclusions as follows:

1.  Two principal sources of free energy are available for amplifying 
a magnetic field in a non-rotating supernova core: the violent convection
occuring within the neutrinosphere (TD93); and within the gain region
in between the neutrinosphere and shock (\cite{T00}).  The equilibrium 
magnetic field in these two convective zones is respectively
$\sim 10^{15}$ and $\sim 10^{14}$ G (eqs. [\ref{brms}] and [\ref{brmsb}]).  
The convection occuring within the neutrinosphere
is a robust consequence of the diffusive transport of heat by neutrinos
(Appendix A).  Even during relatively late fall-back ($\sim 0.1\,M_\odot$
over $\sim 10^3$ s), the neutrino flux is high enough to induce vigorous
convection within the material below the accretion shock.

2.  The supernova core also contains regions which are stable
to Ledoux convection, including the material which straddles the
neutrinosphere(s).  The transport of magnetic fields across these regions
is greatly facilitated by the absorption of electron-type neutrinos
(and anti-neutrinos) on free neutrons (and protons).  The field induces
a chemical potential imbalance between electrons, neutrons, and protons
(cf. Goldreich \& Reisenegger 1992), but the charged-current reactions
force this imbalance toward zero (\S \ref{transport}).  During the main
Kelvin-Helmholtz cooling phase of a nascent neutron star, this process
effectively transports material containing fields stronger than 
$\sim 10^{13}$ G out of the star.

3. During periods of hyper-Eddington accretion, the equililibrium shape 
of a magnetic flux rope which protrudes across the neutrinosphere 
is strongly modified by the hydrostatic pressure of the settling material.
In the case of a flux rope anchored at both ends, the resulting
buoyancy force leads to a critical separation between the ends of
the flux rope, above which the rope expands to very large radius
(\S \ref{secfour}).  This mechanism provides a means of feeding large
amounts of magnetic flux from a dynamo operating within the neutrinosphere,
to the neutrino-heated `gain region' outside the neutron core.
The interaction of this field with the accretion flow has been 
investigated elsewhere (\cite{T00}).

Claims that late fallback will tend to bury a magnetic field (e.g.
Geppert, Page, \& Zannias 1999) should, as a result, be treated with caution.
The full implications of these physical processes for pulsar magnetism
(and perhaps also the supernova mechanism) can only be gleaned by
incorporating them into full numerical simulations of the supernova collapse.
The behavior of a centrifugally-supported accretion flow is, in 
particular, beyond the scope of this paper.  Nonetheless, several 
qualitatitive conclusions will be summarized here, in the following sections.

\subsection{Principal Phase of Dynamo Amplification}

Both the gain region below the accretion shock and the interior
of the neutron star are strongly convective.  The ratio of
convective kinetic energy to gravitational binding energy
is a hundred times larger in the proto-neutron star phase, than during 
any previous convective phase driven by nuclear burning (TD93).  It is even
larger in the convective gain region below an accretion shock 
(\cite{T00}).  Since the ratio of magnetic energy to gravitational 
binding energy is approximately conserved during gravitational collapse, 
these violent fluid motions during the supernova event are expected to be the
dominant source of free energy for pulsar magnetism.  
The relative importance of the two convective regions for the
surface fields of radio pulsars, depends in part on the amount of fallback
following the success of the supernova shock.

Even during late infall ($10^3-10^4$ s after collapse) the settling flow
is heated sufficiently to become violently convective.
As reviewed in \S \ref{dynbal}, the settling speed scales as 
$V(R) \propto \dot M^{3/5}$ when cooling is dominated by $e^\pm$ captures
on free protons and neutrons, whereas the convective speed decreases 
more slowly, $V_{\rm con}(R) \propto \dot M^{1/3}$ (at fixed radius $R$).
The equipartition magnetic field within the flow scales as 
$\langle B^2\rangle^{1/2} \propto \dot M^{8/15}$ (eq. [\ref{eqbscale}]).

\subsection{Prompt Convective Dynamo and the Surface Magnetic Fields 
of Young Neutron Stars}\label{prompt}

The bulk of the crust of a neutron star is derived from material
that is either processed through the convective shell below the
accretion shock; or rises buoyantly from the deep interior of the
neutron core through the convectively stable layer that straddles the
$\nu$-sphere.  The equilibrium flux density (\ref{brmsb}) in the outer
convective shell, scaled to the density of the neutron star crust, is
\be 
B_{\rm crust} \sim \left({\rho_{\rm sat}\over\rho_2}\right)^{2/3}\,
\langle B^2\rangle^{1/2},
\ee 
or equivalently, 
\be 
B_{\rm crust} \sim
2\times 10^{17}\, \Bigl(\varepsilon_B {\cal R}\Bigr)^{1/2}\,
\left({\dot M\over M_\odot~{\rm s}^{-1}}\right)^{-1/6}\,
\left({R_{\rm sh}\over 100~{\rm km}}\right)^{-5/12}\, \left({{\cal
L}_{52}\over 4}\right)^{1/2}\, \left({M_{\rm core}\over
1.4~M_\odot}\right)^{13/12}\;\;\;\;\;\;{\rm G}.  
\ee  
Here $\rho_2$ is
the post shock density (eq. [\ref{rhotwo}]).  This field can be
expected to be coherent on the small scale 
\be 
\ell_B \sim
\left({\rho_2\over\rho_{\rm sat}}\right)^{1/3}\,\ell_P(R_{\rm sh}) =
0.5\,\left({R_{\rm sh}\over 100~{\rm km}}\right)^{-1/2}\, \left({\dot
M\over 1~M_\odot\,{\rm s^{-1}}}\right)^{1/3}\, \left({{\cal R}\over
10}\right)^{1/3}\, \left({M_{\rm core}\over
1.4~M_\odot}\right)^{-1/6}\;\;\;\;\;\;{\rm km}.  
\ee 
This leads to a net dipole field 
\be 
B_{\rm dipole} \sim \langle
B^2\rangle^{1/2}\,\left({\ell_B^2\over 4\pi R_\star^2}\right)^{1/2},
\ee 
namely 
\be\label{bconval}
 B_{\rm dipole} = 3\times
10^{15}\,\varepsilon_{B\,-1}^{1/2}\, \left({\dot M\over M_\odot~{\rm
s}^{-1}}\right)^{1/6}\, \left({R_{\rm sh}\over 100~{\rm
km}}\right)^{-11/12}\, \left({{\cal L}_{52}\over 4}\right)^{1/2}\,
\left({{\cal R}\over 10}\right)^{5/6} \left({M_{\rm core}\over
1.4~M_\odot}\right)^{11/12}\;\;\;\;\;\;{\rm G}.  
\ee 
A further multiplicative factor of $(\ell_B/R_{\rm NS})^{1/2}
\sim (\rho_2/\rho_{\rm sat})^{1/6}$ must be included if one also 
averages the dipole field through many radial shells of thickness $\ell_B$.
This estimate applies to the field generated by convective motions
in the outer part of the gain region.  

Material that rises buoyantly past the $\nu$-sphere must also be strongly
magnetized.  The strong neutrino flux aids buoyancy forces, but the
magnetized fluid will reach the surface of the star during the prompt
Kevlin-Helmholtz phase, only if $B > 10^{13}$ G (eq. [\ref{bmin}]).  
Weaker magnetic
fields that are more characteristic of the dipole fields of ordinary
radio pulsars are not able to reach the neutrinosphere, and according
to our current understanding of transport through the crust will
remain buried for $\sim 10^8$ yr or longer (e.g.  Goldreich \&
Reisenegger 1992).  This suggests that a prompt convective dynamo
can explain the observed magnetic dipole fields of radio pulsars
only if the dynamo operates stochastically -- generating a true dipole
component through the incoherent superposition of stronger, small 
scale fields (TD93).

\subsection{Effects of Late Fallback}\label{fallback}

Even after the shock is pushed to large radius, continuing fallback
(driven by the formation of a reverse shock) can keep the
hydrostatic pressure at the surface of the nascent neutron star well
above the surface value of $B^2/8\pi$ inferred for the dipole fields
of young radio pulsars.  For example, the hydrostatic pressure of
a spherical shell of mass $\Delta M$ corresponds to a field strength
\be
B = \left({2 GM_{\rm NS}\Delta M\over R_{\rm NS}^4}\right)^{1/2}
= 3\times 10^{17}\,\left({\Delta M\over 0.1\,M_\odot}\right)^{1/2}\,
\left({M_{\rm NS}\over 1.4\,M_\odot}\right)^{1/2}\,
\left({R_{\rm NS}\over 10~{\rm km}}\right)^{-2}\;\;\;\;\;\;{\rm G}.
\ee
We focus here on the consequences of {\it spherical} accretion.  
(The behavior of a centrifugally supported flow may be substantially 
different, because it may not be able to maintain the large central 
cusp in the temperature and density needed to effect rapid neutrino 
cooling.)

Consider first the equipartition magnetic field which is maintained
by the convective motions within the material settling onto the neutron
star, driven by neutrino heating.  At low accretion rates, the
accretion shock lies at a large radius, and so it is appropriate
to focus on a fixed radius $R \la 100$ km, where 
$\langle B^2\rangle^{1/2} \propto \dot M^{8/15}$ (eq. [\ref{eqbscale}]).  
Allowing for an additional factor of $\sim (\rho_{\rm sat}/\rho)^{1/6}$ 
-- relating the net dipole field to the equipartition field in the 
settling flow (\S \ref{prompt}) -- one obtains the relation
\be\label{bdipolescale}
B_{\rm dipole} \propto \dot M^{7/15}
\ee
between the dipole field generated by convective fallback, and the
rate of mass accretion.  The estimate (\ref{bconval}) then reduces to
$B_{\rm dipole} \sim 10^{13}$ G at an accretion rate $\dot M = 10^{-4}\,
M_\odot$ s$^{-1}$.

Next, we consider how the continuing flux of electron-type neutrinos
will promote the buoyancy of magnetic fields which are anchored below
the final layer of accreted material.
We start with eq. (\ref{uneu}) for the radial drift speed of a
(horizontal) magnetic flux rope through an atmosphere of non-degenerate
nucleons, given a heating rate (\ref{theat}) by charged-current
absorption of $\nu_e$ and $\bar\nu_e$.  At an intermediate accretion 
rate $10^{-1} \ga \dot M \ga 10^{-4}$, the gravitational binding energy 
is carried off through $e^\pm$ capture on free nucleons in the settling
flow (\cite{T00}), as compared with $e^\pm$ annihilation at lower
accretion rates (Chevalier 1989).  The associated luminosity is directly
related to $\dot M$ by eq. (\ref{lnulate}).
Making use of the scalings (\ref{rhotscal}) of density and temperature
with $\dot M$ (at fixed radius within the settling flow), one finds that the
nucleons become less degenerate as $\dot M$ decreases.  Magnetic flux 
poking out of the nascent neutron star will, as the result of buoyancy 
forces, carry with it a slightly lower mass density and maintain 
a similar degree of degeneracy to the surrounding flow.  

It is now straightforward to derive the critical $\beta_P = B^2/8\pi nT$
above which a parcel of magnetized fluid will rise buoyantly over
the infall time $\tau_{\rm infall} \sim 10^3$ s.  One obtains
\be\label{betapcrit}
\beta_P \ga 1.6 {\tau_{\rm heat}\over \tau_{\rm infall}}
= 1\times 10^{-5}\,\left({\dot M\over 10^{-4}\,M_\odot {\rm s^{-1}}}\right)^{-11/10}\,\left({\tau_{\rm infall}\over 10^3~{\rm s}}\right)^{-1}\,
\left({R_{\rm NS}\over 10~{\rm km}}\right)^3\,
\left({M_{\rm NS}\over 1.4\,M_\odot}\right)^{-1}
\ee
from equations (\ref{theat}) and (\ref{uneu}).
The tiny numerical factor shows that the surface magnetic field
almost certainly cannot be shielded diamagnetically by the accreted
material ($\Delta M \sim \dot M\tau_{\rm infall}$), because in that
circumstance $\beta_P \ga 1$.   Burial is possible only if the 
pre-existing magnetic field undergoes partial turbulent mixing with 
a fraction of the accreted material, so as to force $\beta_P$ below 
the value (\ref{betapcrit}).

\subsection{Origin of the Low-order Magnetic Multipoles of Neutron Stars}

At least two distinct types of hydromagnetic dynamos can give rise to
the dipole magnetic fields of neutron stars.  The dipole
field can either be amplified by fluid stresses that are coherent
over a substantial fraction of the stellar radius (e.g. the classical
$\alpha-\Omega$ dynamo driven by a combination of rapid differential
rotation and convection);  or it can be built up by much smaller
units that themselves are more strongly magnetized.  In the convective
interior of a neutron star, the first type of dynamo requires very rapid
rotation, $P \lsim \ell_P/V_{\rm con} \sim 3$ ms, and has been conjectured
to result in magnetars with dipole fields in excess of $10^{14}$ G.
The second type of stochastic dynamo has been conjectured to generate
the small scale, intranetwork magnetic field of the Sun (TD93;
\cite{DYR}), but is ineffective at generating low order magnetic
multipoles in a main sequence star.  For example, the pressure scale 
height
at the Solar photosphere is only $\sim 10^{-4}$ of the Solar radius,
which implies that the dipole flux density resulting from the incoherent
superposition of the small scale dipoles is suppressed by
a factor $\sim (2\pi R_\odot^2/\ell_P^2)^{-1/2} \sim 10^{-4}$. 
By contrast, $\ell_P \sim R_{\rm ns}/30$ at the neutrinosphere of a
newly formed neutron star,  during the last stages of Kelvin-Helmholtz
cooling.  The corresponding suppression factor is $\sim 10^{-2}$.
This can comfortably accommodate the dipole fields of young radio 
pulsars,
which characteristically lie in the range $\sim 3\times 10^{11} - 
10^{13}$ G.

A similar stochastic dynamo may operate in the convective accretion 
flow onto a nascent neutron star.  During {\it prompt} fallback,
within $\sim 1$ s after collapse, this mechanism could
generate a dipole field as large as $\sim 10^{14}-10^{15}$ Gauss
(eq. [\ref{bconval}]).  However, the equipartition magnetic field
in the convective settling flow becomes weaker with increasing fallback time
(and decreasing $\dot M$; eq. [\ref{bdipolescale}]).
This means that variability in the amount of fallback is an
alternative explanation for the presence of a range of dipole fields 
in young neutron stars -- with magnetars representing those objects 
with the smallest amount of late fallback.  It is interesting to note 
that the current (uncertain) estimates of the mass accreted during 
late fallback ($\Delta M \sim 0.03-0.1\,M_\odot$; e.g. Fryer, Colgate, 
\& Pinto 1999) typically exceed the mass of the rigid crust in model 
neutron stars ($\sim 0.02\,M_\odot$).  As a result, the field generated
during fallback will tend to be anchored in the neutron star core.

\subsection{Observational Tests of the Dynamo Mechanism}

Given the extreme difficulty of inferring the initial spins of
radio pulsars, models of the dynamo origin of pulsar magnetic fields
are most tightly constrained by searches for correlations between
${\bf B}$ and ${\bf\Omega}$.  The stochastic dynamo does not depend
in any way on rotation, and so predicts that these two vectors are
randomly oriented with respect to each other.  By contrast, the magnetic
field formed by a large-scale helical dynamo in a rapidly rotating
neutron star should be strongly correlated with the rotation axis. 
In addition, a neutron star containing a toroidal magnetic field stronger than
$\sim 10^{14}$ G (and an external magnetic dipole that is approximately 
aligned with the symmetry axis of the internal field) has a tendency
to reach an alignment between ${\bf\Omega}$ and the external magnetic moment
(cf. Goldreich 1971).  This effect is more likely to induce alignment 
in the Soft Gamma Repeaters and Anomalous X-ray Pulsars, than in young 
radio pulsars (Duncan, Li \& Thompson 1994).

\subsection{Relaxation of Small-Scale Magnetic Fields}

A small scale magnetic field of strength $10^{14}-10^{15}$ G in
radio pulsars would be manifested through sudden fractures of the
neutron star crust.  This phenomenon could possibly play a role in
triggering glitches (TD93) by intermittently heating the crust, or
inducing rapid translations of the Coulomb lattice with with respect
to the neutron superfluid.  Nonetheless, the absence of X-ray bursts
associated with radio pulsars, combined with the seismic quiescence of
old radio pulsars, suggests that if such a field is present it must be
buried fairly deeply.  Either a radio pulsar must have accreted
upwards of $\sim 10^{-2}\,M_\odot$ of {\it weakly} magnetized material 
($B \sim 10^{12}-10^{13}$ G) following the brief dynamo epoch 
(eq. [\ref{bdipolescale}]), or the high order magnetic multipoles produced 
by convection were effectively smoothed out.  

The stability of higher magnetic multipoles depends on the ability
of the gravitationally stratified fluid to undergo fully three-dimensional
motions (TD93).  Stable stratification, which forces the fluid 
to move along the equipotential surfaces of the star, may allow more 
complicated magnetic topologies.  However, we have shown that
the absorption and emission of electron-type neutrinos facilitates
the buoyant motion of magnetic fields as weak as $\sim 10^{13}-10^{14}$ G
(during the prompt Kelvin-Helmholtz phase of a proto-neutron star),
and so will facilitate the reconnection and unwinding of stronger
fields.


\vskip 0.4in
\centerline{ACKNOWLEDGEMENTS}

We gratefully acknowledge conversations with Phil Arras, Leon Mestel, Ewald
Mueller, and Mal Ruderman.  Financial support has been provided to 
both authors by the NSERC of Canada; and to C.T. by the Sloan Foundation.

\vskip 0.4in
\appendix
\section{Convection in Young Neutron Stars}

In this appendix we calculate the temperature gradient required to
trigger convection in the presence of the stabilizing composition
gradient found in a cooling neutron star. This
gradient is derived below and given in eqn. (\ref{convective}) in the
main text. We compare the result to that of Lattimer \& Mazurek
(1981), who find that lepton-fraction gradients tend to be
stablizing at low lepton fraction, and Reisenegger \& Goldreich
(1992), who find that the $Y_e$-gradient in a cold neutron star at
$\beta$-equilibrium is absolutely stabilizing.  The critical value of
$Y_l$, below which the lepton gradient is stabilizing, increases at
high temperatures and low densities, a result embodied in
eqn. (\ref{sderivb}) below.  Although more detailed numerical cooling
models are now available (e.g. \cite{pons}), the following arguments
are useful for the physical insight they provide.

We are interested in the response of a neutron star to displacements
of fluid elements. The temperature and density of the stellar material
vary with radius; moving material radially will result in buoyancy
forces which can either oppose or enhance the force initiating the
displacement, depending on the thermodynamic state of the star. We
assume that the star is in hydrostatic equilibrium, with equilibrium
density and pressure gradients $d\rho_0/dR$ and $dP_0/dR$, related by
\be
{dP_0\over dR} = -\rho_0 g
\ee
and an equation of state
\be
P = P(\rho, T, Y_e, Y_{\nu_e}).
\ee
The density and temperature are assumed to be high enough that the
stellar material is composed entirely of neutrons, protons, electrons,
and neutrinos.\footnote{But not so high that $\mu$ and $\tau$ 
leptons, pions, or strange particles are generated in significant abundances.}
The relative proportions of these species are
related to the total nucleon density\footnote{We ignore the electron
mass as well as the small mass difference $m_n-m_p$.} $n_b = \rho/m_n$ by the
parameters $Y_e = (n_{e^-}-n_{e^+})/n_b$, $Y_p = n_p/n_b = Y_e$
(from overall electric charge neutrality), 
$Y_{\nu_e} = (n_{\nu_e}-n_{\bar\nu_e})/n_b$, and $Y_n = n_n/n_b =
1-Y_e$.  The total lepton fraction is
\be
Y_l = Y_e + Y_{\nu_e}.
\ee

We model each species of particle as an almost degenerate,
ideal fermi gas with chemical potential $\mu_i$.  The background
configuration is assumed to be in chemical equilibrium:
\be\label{chem}
\mu_n + \mu_{\nu_e} = \mu_p + \mu_e.
\ee
In a hot neutron star ($T \ga 10$ MeV), which is optically thick to all
the neutrino species, a displaced fluid element can be assumed to
remain in chemical equilibrium.  At very high temperatures, the 
lepton fraction can also be assume to be frozen into a buoyant
plume of material, $Y_l \simeq$ constant.  The convective efficiency
is significantly reduced once the optical depth $\tau_{\nu_e}$ to electron
neutrinos across a pressure scale height drops to a value
$\sim c/V_{\rm con} \sim 300$, where $V_{\rm con} \sim 10^8$ cm s$^{-1}$ 
is the speed of the convective overturns (cf. \cite{Mezz98}).  
The corresponding mean free path to $\nu_e + n\rightarrow p + e^-$
is $\lambda \sim 10^3$ cm, which is reached at a temperature 
$k_BT \sim 15$ MeV at nuclear saturation density  
($n_b = n_{\rm sat} = 1.6\times 10^{38}$ cm$^{-3}$; cf.
Fig. 7 of Reddy et al. 1998).  In fact, as we now show, this temperature
is comparable to the value below which a purely radiative flux of
neutrinos can be maintained without inducing convective instability.

\subsection{Convective Instability}

We first review the derivation of the buoyancy force and the 
Brunt-V\"ais\"al\"a frequency $N$ associated with the displacement
of a fluid element from its equilibrium position.  After
a displacement through a distance $\xi$ in the radial
direction, the density of a fluid element becomes
\be 
\rho_d(R+\xi)=\rho(R)+\left({\partial\rho\over\partial R}\right)_{S,Y_l}\xi.
\ee 
The subscripts indicate that the entropy and composition are
held fixed in calculating the derivative.

The buoyancy force arises because this density differs from the
background density at $r+\xi$:
\be 
\rho_0(R+\xi)=\rho(R)+{d\rho_0\over dR}\xi.
\ee 
The (Eulerian) density difference is
\be \label{euler}
\delta\rho\equiv\rho_d(R+\xi)-\rho_o(R+\xi)\approx
\left({d\rho\over dR}\right)_{S,\by}
\xi
-{d\rho_0\over dR}\xi,
\ee 
resulting in a buoyancy force
\be \label{buoyancy}
F_B=-g\delta\rho=-g\left[\left({d\rho\over dR}\right)_{S,\by}
-{d\rho_0\over dR}\right]\xi
\ee 
Suppose we move a bit of fluid upward, so that $\xi>0$. If the density
of the displaced fluid element is less than that of the surrounding
material ($\delta\rho<0$) then the term inside the square brackets in
eqn. (\ref{buoyancy}) is negative.  The buoyancy force is upward, in
the direction of the original displacement, and the star is unstable
to convection. 

Young neutron stars have both entropy and composition
gradients, so it is convenient to express the buoyancy force and $N$
in terms of those gradients. Noting that
$\rho(R)=\rho\left[P(R),S(R),Y_l(R)\right]$ we find 
\be  
\left({d\rho\over dR}\right)_{S,Y_l}=
\left({\partial\rho\over\partial P}\right)_{S,Y_l}{dP_0\over dR}
\ee 
and
\be  \label{rho_S_Y}
\left[
\left({\partial\rho\over\partial P}\right)_{S,Y}{dP_0\over dR}
-{d\rho_0\over dR}\right]=
-\left[
\left({\partial\rho\over\partial S}\right)_{P,Y}{dS_0\over dR}
+\left({\partial\rho\over\partial Y}\right)_{P,S}{dY_{l,0}\over dR}
\right].
\ee 

The condition for convective instability becomes
\be  
\left[
\left({\partial \rho\over\partial S}\right)_{P,Y_l}{dS_0\over dR}+
\left({\partial\rho \over\partial Y_l}\right){dY_{l,0}\over dR}
\right]>0
\ee 
Since $\left({\partial \rho\over\partial S}\right)_{P,Y_l}<0$, 
the condition for triggering convection is
\be  
{dS_0\over dR} + {(\partial\rho/\partial Y_l)_{P,S}\over 
(\partial\rho/\partial S)_{P,Y_l}}{dY_{l,0}\over dR} 
= {dS_0\over dR} - \left({\partial S\over \partial Y_l}\right)_{P,\rho}\,
{dY_{l,0}\over dR} < 0,
\ee 
which is eqn. (\ref{ledcon}) in the main text.  

\subsection{Influence of a Compositional Gradient}

The sign of the thermodynamic derivative $(\partial S/\partial Y_l)_{P,\rho}$
determines whether the lepton gradient is stabilizing or de-stabilizing.  
Note that $dY_l/dR < 0$ is quickly established by passive neutrino diffusion
in a proto-neutron star.  Evaluation of the derivative is greatly
simplified when $Y_\nu \ll Y_e$, so that one can approximate 
$\partial/\partial Y_l \simeq \partial/\partial Y_e$.  

We idealize the neutrons and protons as non-relativistic Fermi gases.
It will prove convenient to introduce a characteristic Fermi momentum
\be\label{pfval} 
p_F\equiv\hbar(3\pi^2n_b)^{1/3} = 330\,
\left({n_b\over n_{\rm sat}}\right)^{1/3}\;\;\;\;\;\;{\rm MeV/c}.
\ee 
Then the Fermi energies of the neutrons and protons can be written as
\be  
\eps_{p,n}\approx{\hbar^2\over 2m_n}(3\pi^2n_{p,n})^{2/3} =
Y_{p,n}^{2/3}\,{p_F^2\over 2m_n}.
\ee  
Here $\hbar$ is Planck's constant. The neutrons are in fact just 
becoming degenerate at $k_BT\approx 40$ MeV, which is the peak 
temperature generated immediately after the collapse (\cite{pons}); it is the
associated rise in pressure that halts the collapse. The protons are
less dense, and so they are only marginally degenerate.  Their
contribution to the entropy is smaller but significant.  The analysis
is simplified by treating the protons as an almost degenerate gas,
but it should be kept in mind that this approximation is only marginally
valid.

The electrons and electron neutrinos are, by contrast, highly
relativistic (cf. equation [\ref{pfval}]).
Their Fermi energies are
\be 
\eps_e\approx\hbar c(3\pi^2n_e)^{1/3} = 
Y_e^{1/3}\,p_F c.
\ee 
and
\be 
\eps_{\nu_e}\approx\hbar c(6\pi^2n_{\nu_e})^{1/3} = 
(2Y_{\nu_e})^{1/3}\,p_F c.
\ee 
Notice that 
\be
{\eps_{\nu_e}\over\eps_e} = \left({2Y_{\nu_e}\over Y_e}\right)^{1/3}
\ee
is not necessarily small when $Y_{\nu_e} \ll Y_e$.  We will
{\it not} neglect terms of order $\eps_{\nu_e}/\eps_e$
even while dropping terms of order $Y_{\nu_e}/Y_e$.

At finite temperature, the chemical potentials are
\be 
\mu_{p,n}\approx\eps_{p,n}\left[1-{\pi^2\over12}
	\left({\kb T\over\eps_{p,n}}\right)^2\right],
\ee 
and
\be 
\mu_{e,\nu}\approx\eps_{e,\nu}\left[1-{\pi^2\over3}
	\left({\kb T\over\eps_{e,\nu}}\right)^2\right],
\ee 
and the entropies (per baryon) are
\be \label{baryonentropy} 
{S_{p,n}\over\kb}={\pi^2\over2}Y_{p,n}\,\left({\kb T\over \eps_{p,n}}\right)
\ee 
and
\be 
{S_{e,\nu}\over\kb}=\pi^2Y_{e,\nu}\,\left({\kb T\over \eps_{e,\nu}}\right).
\ee 
The pressures can be approximated as
\be \label{bpressure}
P_{p,n}={2\over5}n_{p,n}\eps_{p,n} + {1\over 3}n_b S_{p,n}T
\ee 
and
\be \label{epressure}
P_{e,\nu}={1\over4}n_{e,\nu}\eps_{e,\nu} + {1\over 6}n_b S_{e,\nu_e} T.
\ee 

The key thermodynamic derivative can be expanded as
\be\label{sderiv} 
\left({\partial S\over\partial Y_l}\right)_{P,\rho} \simeq
\left({\partial S\over\partial Y_e}\right)_{P,\rho} =
\left({\partial S\over\partial Y_e}\right)_{\rho,T} +
\left({\partial S\over\partial T}\right)_{\rho,Y_e}
\left({\partial T\over\partial Y_e}\right)_{P,\rho}.
\ee 
To evaluate this expression, we need 
\be \label{dsdT_rho} 
\left({\partial S\over\partial T}\right)_{\rho,Y_e}={S\over T}.
\ee 
and
\be 
\left({\partial T\over \partial Y_e}\right)_{P,\rho} =
-{(\partial P/\partial Y_e)_{\rho, T}\over (\partial P/\partial T)_{\rho,Y_e}}.
\ee 
The derivative $(\partial P/\partial Y_e)_{T,\rho}$ is dominated
by the (zero-temperature) degeneracy pressure 
\be
P_0 \equiv 
{2\over 5}n_n\eps_n + {2\over 5}n_p\eps_p + {1\over 4}n_e\eps_e
+ {1\over 4}n_{\nu_e}\eps_{\nu_e},
\ee
so that
\be\label{p0deriv}
\left({\partial P_0\over\partial Y_e}\right)_\rho
= {1\over 3}n_b(2\epsilon_p-2\epsilon_n+\epsilon_e) +
{1\over 3}n_b\epsilon_{\nu_e}\,
\left({\partial Y_{\nu_e}\over \partial Y_e}\right)_\rho
\ee
Neglecting the last term on the RHS for the moment, and invoking the 
$\beta$-equilibrium condition $\eps_p+\eps_e=\eps_n+\eps_\nu$, one finds
\be 
\left({\partial P_0\over\partial Y_e}\right)_\rho={n_b\over3}
\left(2\eps_\nu-\eps_e\right).
\ee 
However, the thermal contribution to the pressure cannot be neglected,
\be
\left({\partial P\over\partial Y_e}\right)_{\rho,T} = 
\left({\partial P_0\over\partial Y_e}\right)_\rho +
{n_b T\over 3}{\partial\over \partial Y_e}\left[S_p + S_n +
{1\over 2}\left(S_e + S_{\nu_e}\right)\right],
\ee
as it rescales the first term on the RHS of eq. (\ref{sderiv}).
Combining 
\be \label{dPdT_rho}
\left({\partial P\over\partial T}\right)_{\rho,Y_e}={2\over3}n_b
\left(S_p +S_n + {1\over 2}S_e + {1\over 2}S_{\nu_e}\right),
\ee 
with the above equations, one finds
$$
\left({\partial S\over\partial Y_e}\right)_{P,\rho} =
{1\over 2}\left[\left({\partial S_n\over\partial Y_e}\right)_{\rho,T} +
\left({\partial S_p\over\partial Y_e}\right)_{\rho,T}\right]\,
\left[{S_p + S_n\over S_p + S_n + (S_e + S_{\nu_e})/2}\right] +
$$
\be\label{S_Y} 
+{3\over 4}\left[\left({\partial S_e\over\partial Y_e}\right)_{\rho,T} +
\left({\partial S_{\nu_e}\over\partial Y_e}\right)_{\rho,T}\right]\,
\left[{S_p + S_n+(S_e+S_{\nu_e})/3\over S_p + S_n + (S_e + S_{\nu_e})/2}\right]
- \left[{S\over S_p + S_n + (S_e +S_{\nu_e})/2}\right]
{2\eps_{\nu_e}-\eps_e\over 2T}
\ee 
Neglecting the entropy of the electrons and neutrinos compared
with the nucleons, this expression simplifies to 
\be\label{sderivb} 
\left({\partial S\over\partial Y_e}\right)_{P,\rho} =
{1\over 2}\left({\partial S\over\partial Y_e}\right)_{\rho,T}
- {2\eps_{\nu_e}-\eps_e\over 2T}
\ee 
The variation of the entropy with lepton number is, in the same
approximation,
\be \label{sye}
\left({\partial S\over\partial Y_e}\right)_{\rho,T} = 
{1\over3}\left({S_p\over Y_e}-{S_n\over 1-Y_e}\right)
= {\pi^2\over 3}k_B\left({k_BT\over\eps_e}\right)\,
\left({m_nc^2\over\eps_e}\right)\,
\left[1-\left({Y_e\over 1-Y_e}\right)^{2/3}\right].
\ee 

A fuller evaluation of these derivatives, including the contributions
from the electrons and electron neutrinos, requires knowing
the quantity $(\partial Y_{\nu_e}/\partial Y_e)_{\rho,T}$, 
which can be determined directly from the condition (\ref{chem})
of chemical equilibrium:
$(\partial Y_{\nu_e}/\partial Y_e)_{\rho,T} \simeq (Y_{\nu_e}/Y_e)^{2/3}$
when $Y_{\nu_e} \ll Y_e$.  For example, substitution of this expression on
the RHS of eq. (\ref{p0deriv}) shows that the derivative of the
electron neutrino pressure is of order $Y_{\nu_e}/Y_e$, and can indeed
be neglected.

The compositional gradient ($dY_l/dR < 0$) tends to suppress
convective instability if $(\partial S/\partial Y_l)_{P,\rho} > 0$,
and is otherwise de-stabilizing.  Both terms in eq. (\ref{sderivb})
are positive if $2\eps_\nu < \eps_e$, that is, if $Y_{\nu_e} < {1\over 16}Y_e$.
However, immediately after the collapse $\eps_\nu\approx\eps_e$, and
so the composition gradient is potentially de-stabilizing.  
The sign of $(\partial S/\partial Y_l)_{P,\rho}$ reverses
at a critical temperature
\be\label{tstarb} 
kT_\star \equiv {\eps_e\over \pi}
\left[{3(2\eps_\nu-\eps_e)\over m_nc^2}\right]^{1/2}\,
\left[1-\left({Y_e\over 1-Y_e}\right)^{2/3}\right]^{-1/2},
\ee 
which we evaluate in the main text.

This expression qualitatively reproduces the domain of 
convective instability in the $S-Y_l$ plane, as plotted in
Figure 1 of Lattimer \& Mazuerk (1981).  Those authors find that low
$Y_l$ and high $T$ tend to suppress compositionally-driven convection. 
They also find that convective instability remains possible at very 
low entropies and lepton fractions, where heavy nuclei can form.  However, that
instability is restricted to densities below nuclear saturation.

\subsection{Triggering Convection}

When the compositional gradient is destabilizing, convection will
ensue. However, when it is stabilizing, convection may still occur 
if the entropy gradient is negative.  Convection occurs if
\be \label{trigger}
{dS_0\over dR}<
\left({\partial S\over\partial Y_l}\right)_{P,\rho}
{dY_{l,0}\over dR}.
\ee 

The gradient of $Y_l$ is generally negative, and we have
already evaluated the thermodynamic derivative 
$(\partial S/\partial Y_l)_{P,\rho}$.  It remains to re-express
the entropy gradient in terms of a convective temperature gradient.
Approximating
\be 
S \simeq S_n + S_p = {p_F m_n k_B^2T\over 3\hbar^3 n_b}\,
\left[Y_e^{1/3} + (1-Y_e)^{1/3}\right],
\ee 
(recall $p_F =\hbar\,(3\pi^2 n_b)^{1/3}$), the left hand side of
eqn. (\ref{trigger}) yields
\be\label{sderivc} 
{d\ln S\over d\ln R} = {d\ln T\over d\ln R} - {2\over 3}
{d\ln \rho\over d\ln R} 
+ 
{
	Y_e^{1/3}\left[\yn^{2/3}-Y_e^{2/3}\right]
\over 
	\yn^{2/3}\left[3\yn^{1/3}+Y_e^{1/3}\right]
}
\,{d\ln Y_e\over d\ln R}.
\ee 
We evaluate the first factor on the left hand side of
eqn. (\ref{trigger}) at low temperatures. This allows us to neglect
the first term in eq. [\ref{sderivb}], yielding
\be\label{sderivd} 
{Y_e\over S}\left({\partial S\over\partial Y_e}\right)_{P,\rho} \simeq
{3Y_e^{4/3}\over Y_e^{1/3}+(1-Y_e)^{1/3}} {\hbar^3 c n_b\over m_n
(k_BT)^2}\,\left({2\mu_{\nu_e}-\mu_e\over2\mu_e}\right).  
\ee 
Using eqs. (\ref{sderivc}) and (\ref{sderivd}) in (\ref{trigger}) leads to 
expression (\ref{convective}) for the convective temperature
derivative $(d\ln T/d\ln R)_{\rm conv}$.  In \S \ref{inside}, we
calculate the radiative temperature gradient associated with passive
neutrino transport, and infer the critical radiative flux that drives
convection.

\subsection{The Brunt-V\"ais\"al\"a frequency \label{brunt}}

The electron fraction decreases outward through the
degenerate core of a cold neutron star.  That this compositional
gradient tends to stabilize against convection was noted by
Lattimer \& Mazurek (1981) and demonstrated in detail by 
Reisenegger \& Goldreich (1992) (in the 
approximation where the neutrons and protons are normal Fermi
fluids).  Here we reproduce the result of Reisenegger \& Goldreich
(1992).

The equation of motion
for our fluid element is
\be 
\rho{d^2\xi\over dt^2}=F_B.
\ee 
This leads to the definition of the Brunt-V\"ais\"al\"a frequency $N$,
\be \label{bruntn} 
N^2\equiv{g\over\rho}\left[\left({d\rho\over dr}\right)_{S,\by}
-{d\rho_0\over dr}\right]=
-{g\over\rho}\left({\partial\rho\over\partial S}\right)_{P,Y}
\left[
{dS_0\over dr}
-\left({\partial S\over\partial Y_l}\right)_{P,\rho}{dY_0\over dr}
\right].
\ee 
We use the latter form in order to check our expressions for the
partial derivatives by comparing to the result of Reisenegger \&
Goldreich (1992), who
evaluate the Brunt in a cold neutron star. The star is assumed contain
no neutrinos ($Y_{\nu_e}=\eps_\nu=0$), to be in chemical equilibrium
($\eps_n\approx\eps_e$), and to have zero temperature and
entropy. Under those conditions the Brunt reduces to
\be  
N^2={g\over\rho_0}\left({\partial\rho\over\partial S}\right)_{P,Y_e}
\left({\partial S\over\partial Y_e}\right)_{P,\rho}{dY_e\over dr}.
\ee 
 From the definition $Y_e=n_e/n_b$ we have 
\be 
{dY_e\over dr}={1\over n_b}{dn_e\over dr}-{Y_e\over n_b}{dn_b\over dr}.
\ee 
 From the equilibrium condition we find
\be 
\alpha n_n^2=n_e,
\ee 
where $\alpha\equiv (\hbar/2mc)^33\pi^2$ is a numerical constant. From
charge neutrality $n_e=n_p$, conservation of baryon number gives
$n_n+n_e=n_b$. Taking the derivative of this expression with respect
to radius, and solving for the electron density gradient, we find
\be \label{electron}
{dn_e\over dr}=2Y_e{dn_b\over dr}+O(Y_e)^2.
\ee 
Combining these results in
\be \label{composition}
{dY_e\over dr}={Y_e\over\rho_0}{d\rho_0\over dr}.
\ee 

Finally we evaluate the variation of density with entropy at fixed
pressure and composition. We have
\be 
\left({\partial S\over\partial\rho}\right)_{P,\by}=
\left({\partial S\over\partial\rho}\right)_{T,\by}-
\left({\partial S\over\partial T}\right)_{\rho,\by}
{\left({\partial P/\partial\rho}\right)_{T,\by}\over
\left({\partial P/\partial T}\right)_{\rho,\by}}.
\ee 

From equations (\ref{baryonentropy}-\ref{epressure}) we find
\be 
\left({\partial S\over\partial \rho}\right)_{T,Y_e}=-{2S\over 3\rho},
\ee 
\be 
\left({\partial P\over\partial \rho}\right)_{T,Y_e}={5P\over 3\rho},
\ee 
and
\be 
\left({\partial P\over\partial T}\right)_{\rho,Y_e}=-{2\over 3}n_bS,
\ee 
where we neglect the leptonic contribution to the entropy. Combining
these with eqn. (\ref{dsdT_rho}) we find
\be 
\left({\partial S\over\partial\rho}\right)_{P,Y_e}=
-\left({2S\over3\rho}\right)
-\left({5S\over2\rho}\right){P\over n_bST}.
\ee 
The second term on the right-hand side of this equation is much larger
than the first: $P/n_bST\approx (4/5\pi^2)(\eps_n/k_bT)^2>>1$. 
To lowest order in $k_bT/\eps_b$
\be \label{rho_S}
\left({\partial\rho\over\partial S}\right)_{P,Y_e}=
-{\pi^2\over2}\left({k_bT\over \eps_n}\right)^2{\rho\over S}
=-{T\over\eps_n}{\rho\over Y_n}.
\ee 
Using equations (\ref{sderivb}), (\ref{composition}), and (\ref{rho_S})
\be  
N^2={g\over\rho_0}
\left(-{T\over\eps_n}{\rho_0\over Y_n}{\eps_n\over2T}\right)
{Y_e\over\rho_0}{d\rho_0\over dr}={Y_e\over 2Y_n}{g\over H},
\ee 
where $H\equiv -\rho/(d\rho_0/dr)$. This is the expression found by
Reisenegger \& Goldreich (1992).

\clearpage
\begin{figure}
\figurenum{1}
\plotone{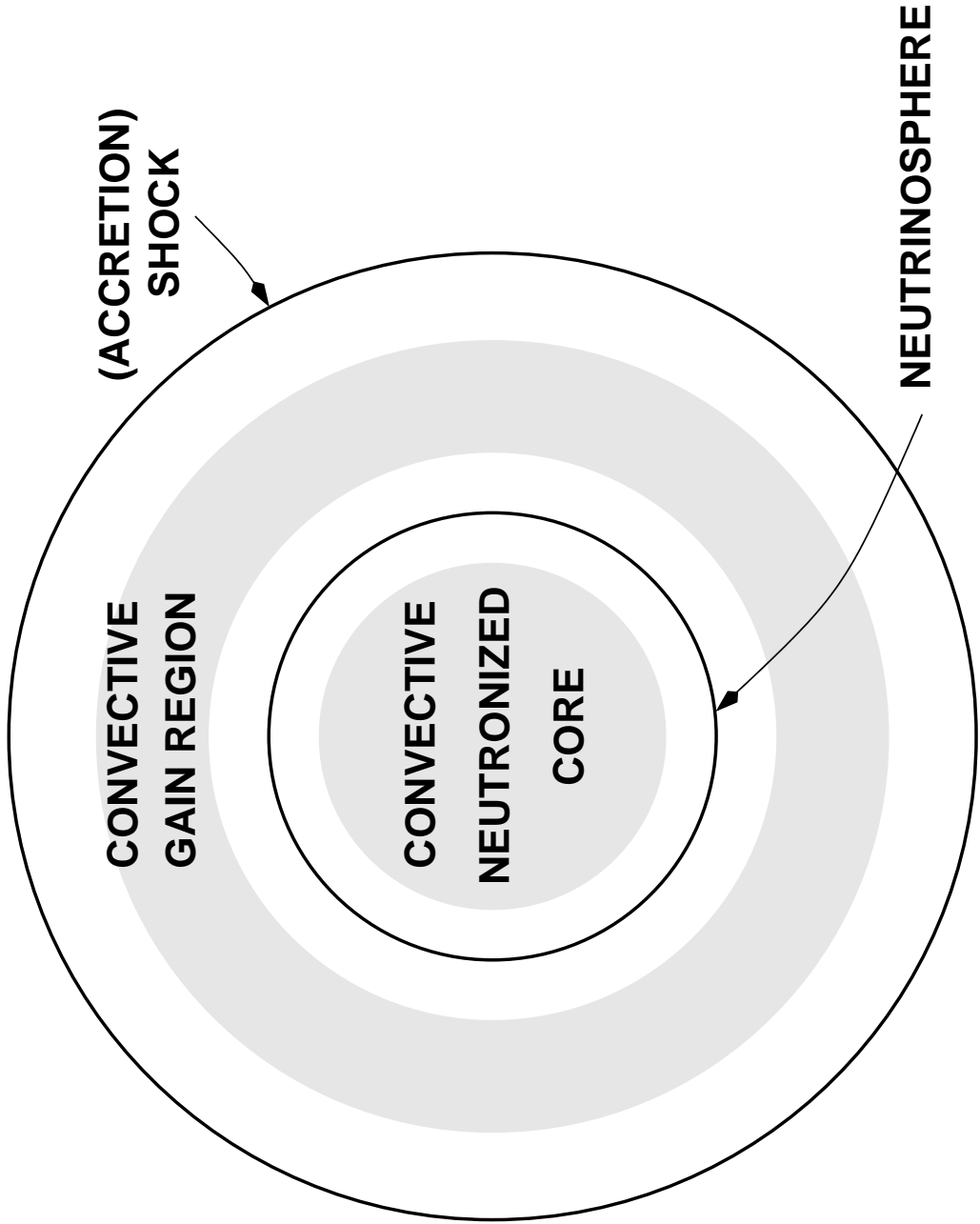}
\caption{
The collapsed core of a massive star develops a violent convective
instability in two distinct regions:  within the central, neutronized
core which is optically thick to neutrinos;  and outside the
neutrino photosphere, within a spherical shell where heating
by the charged-current absorption of $\nu_e$ and $\bar\nu_e$ on
free $n$, $p$ is faster than cooling by captures of electron-positron
pairs.  The region straddling the $\nu$-sphere is, typically,
stable to convection.}
\label{convect}
\end{figure}

\clearpage
\begin{figure}
\figurenum{2}
\plotone{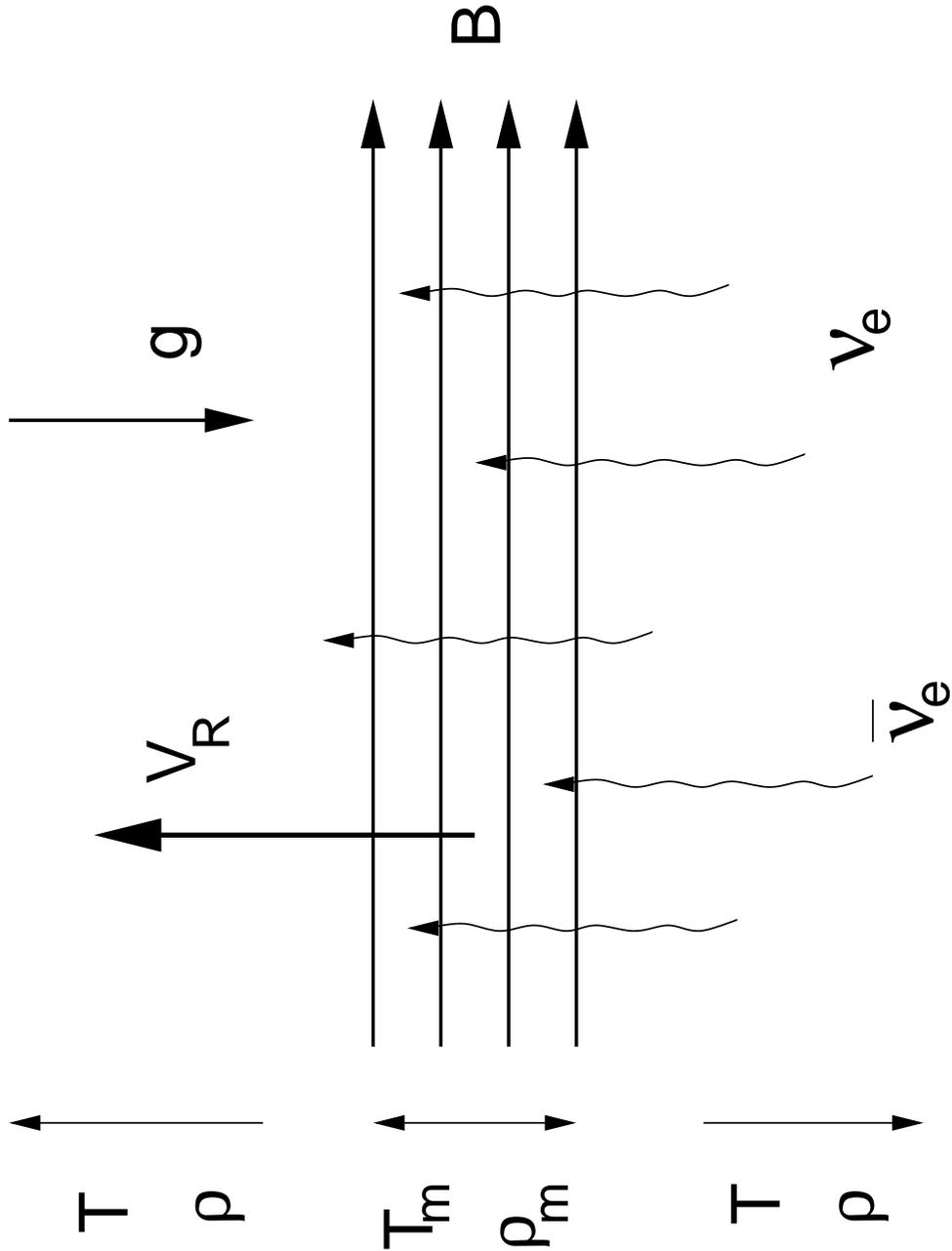}
\caption{
A horizontal bundle of magnetic flux has a lower temperature than
its surrroundings, $T_M < T$, at constant density and pressure.
Electron-type neutrinos, originating outside the bundle, raise
its temperature and allow it to move upward through a stabilizing 
composition (or entropy) gradient.}
\label{buoyant}
\end{figure}

\clearpage
\begin{figure}
\figurenum{3}
\plotone{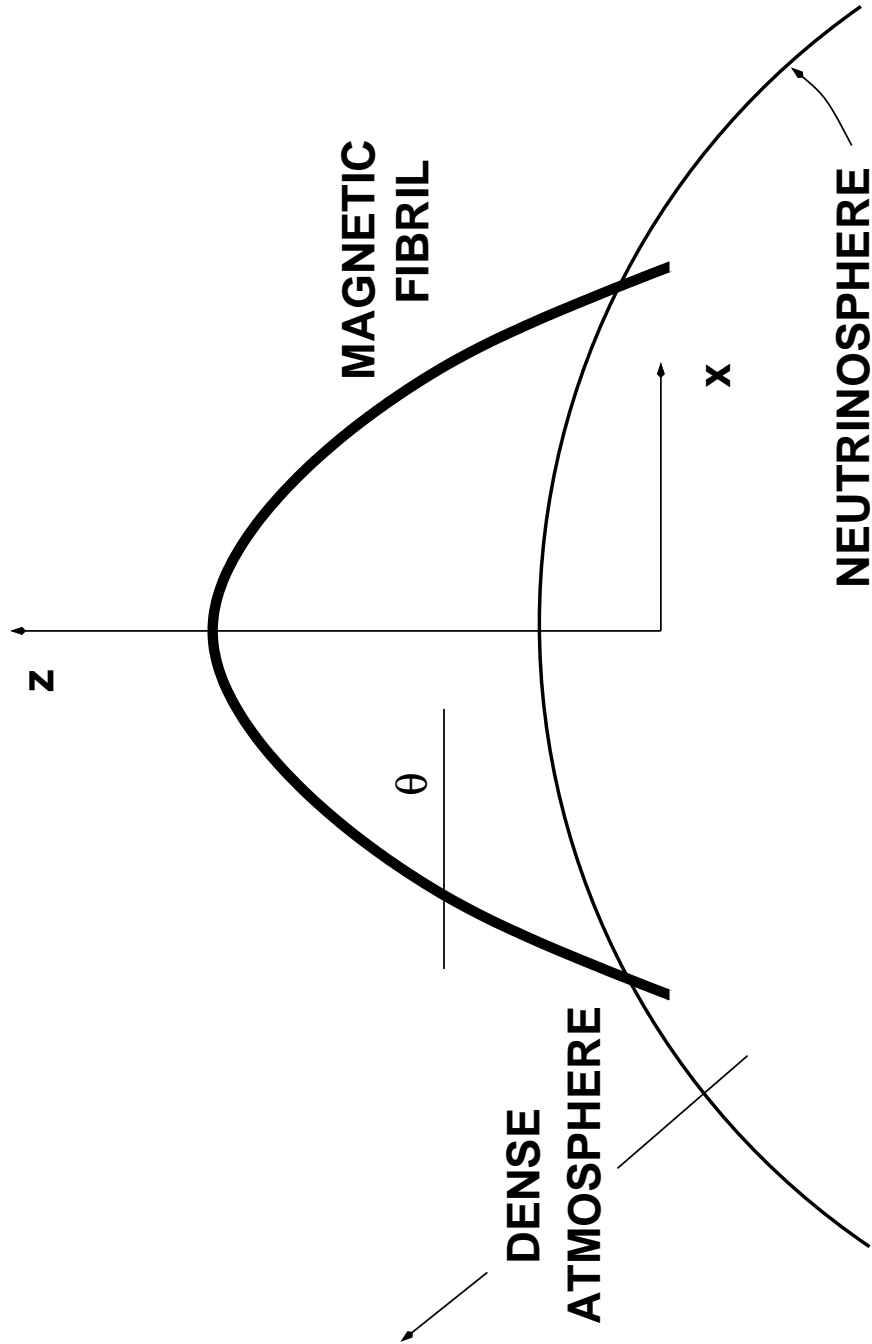}
\caption{
A narrow bundle of magnetic flux reaches an equilibrium configuration
which is a competition between buoyancy and tension forces.  When the
separation between the magnetic footpoints exceeds a critical value,
the equilibrium bundle reaches to infinite height (radius).}
\label{fibril}
\end{figure}

\end{document}